\documentclass[screen, sigconf, nonacm]{acmart}
\usepackage{placeins}
\usepackage{array}
\usepackage{graphicx}
\usepackage{pgfplots}
\usepackage{booktabs}
\usepackage{multirow}
\usepackage{listings}
\usepackage{verbatimbox}
\usepackage{threeparttable}
\usepackage{tablefootnote}
\usepackage{paralist}
\usepackage{xurl}
\usepackage{tcolorbox}
\usepackage{xcolor}
\PassOptionsToPackage{table,dvipsnames}{xcolor}
\usepackage{xspace}
\usepackage{adjustbox}
\usepackage[nameinlink,capitalise]{cleveref}
\usepackage{wasysym} 
\usepackage[inline]{enumitem}
\usepackage{moreenum}
\newlist{paraenum}{enumerate*}{1}
\setlist[paraenum]{label=\emph{(\arabic*)}}
\usepackage{subcaption}
\usepackage{fontawesome}

\usepackage{siunitx}
\usepackage[table]{xcolor}
\usepackage{colortbl}

\definecolor{OIblue}{RGB}{0,114,178}
\definecolor{OIorange}{RGB}{230,159,0}
\definecolor{OIvermillion}{RGB}{213,94,0}

\colorlet{sev1}{OIblue!20}        
\colorlet{sev2}{OIorange!28}      
\colorlet{sev3}{OIvermillion!28}  
\colorlet{sev4}{OIvermillion!45}  
\newcommand{\rowseverity}[1]{%
    \ifdim #1 pt < 10pt
        \rowcolor{sev1}%
    \else\ifdim #1 pt < 30pt
        \rowcolor{sev2}%
    \else\ifdim #1 pt < 50pt
        \rowcolor{sev3}%
    \else
        \rowcolor{sev4}%
    \fi\fi\fi
}

\definecolor{bblue}{rgb}{0.1,0.51,1}
\definecolor{rred}{rgb}{0.83,0.07,0.35}
\definecolor{linkcolor}{rgb}{0.65,0,0}
\definecolor{citecolor}{rgb}{0,0.65,0}
\definecolor{urlcolor}{rgb}{0,0,0.65}
\definecolor{dkgreen}{rgb}{0,0.6,0}
\definecolor{gray}{rgb}{0.5,0.5,0.5}
\definecolor{lightgray}{rgb}{0.8,0.8,0.8}
\definecolor{mauve}{rgb}{0.58,0,0.82}

\definecolor{okColor}{HTML}{4faf4d}
\definecolor{flawedColor}{HTML}{fe0000}
\definecolor{incompleteColor}{HTML}{ffc165}
\definecolor{unclearColor}{HTML}{ff8000}
\definecolor{naColor}{HTML}{387fb7}

\hypersetup{colorlinks=true, linkcolor=linkcolor, urlcolor=urlcolor, citecolor=citecolor}

\lstdefinestyle{codesample}{
    basicstyle=\footnotesize,
    frame=lines,
    language=python,
    aboveskip=3mm,
    belowskip=3mm,
    showstringspaces=false,
    columns=flexible,
    numbers=left,
    numberstyle=\footnotesize\color{black},
    numbersep=5pt,
    morekeywords={False, True, uint8\_t, bool},
    keywordstyle=\color{blue},
    commentstyle=\color{dkgreen},
    stringstyle=\color{mauve},
    breaklines=true,
    breakatwhitespace=true,
    tabsize=2,
    captionpos=b,
    belowskip=-1.75 \baselineskip
}

\pgfplotsset{compat=1.16} 
\usetikzlibrary{calc,positioning,shapes.misc}

\newcommand{\eg}{\textit{e.g.},\xspace}
\newcommand{\ie}{\textit{i.e.},\xspace}

\newcommand{\aref}[1]{\hyperref[#1]{appendix~\ref*{#1}}}

\newcommand{\PrimeProbe}{\textsc{Prime+Probe}\xspace}
\newcommand{\FlushReload}{\textsc{Flush+Reload}\xspace}

\newcommand{\naKeyword}{\textcolor{naColor}{\emph{NA}}}
\newcommand{\flawedKeyword}{\textcolor{flawedColor}{\emph{Flawed}}}
\newcommand{\unclearKeyword}{\textcolor{unclearColor}{\emph{Unclear}}}
\newcommand{\incompleteKeyword}{\textcolor{incompleteColor}{\emph{Incomplete}}}
\newcommand{\okKeyword}{\textcolor{okColor}{\emph{Ok}}}

\newcommand{\flawref}[1]{\hyperref[flaw:#1]{#1}}
\newcommand{\codebookref}[1]{\hyperlink{codebook:#1}{#1}}
\newcommand{\defref}[1]{\hyperref[code_to_flaw:#1]{#1}}

\setdefaultenum{\color{black} (i)}{}{}{}

\begin{document}

\title{Practice Makes (Im)Perfect: 
A Look Back at Benchmarking Practices for Microarchitectural Side-Channel Attacks
}

\author{Iliana Fayolle}
\email{iliana.fayolle@inria.fr}
\orcid{0009-0000-6690-529X}
\affiliation{%
  \institution{Univ. Lille, CNRS, Inria}
  \city{Lille}
  \country{France}
}

\author{Antoine Geimer}
\email{antoine.geimer@inria.fr}
\orcid{0009-0004-0704-8717}
\affiliation{%
  \institution{Univ. Lille, CNRS, Inria}
  \city{Lille}
  \country{France}
}

\author{Daniel De Almeida Braga}
\email{daniel.de-almeida-braga@irisa.fr}
\orcid{0000-0001-5083-5434}
\affiliation{%
  \institution{Univ. Rennes, CNRS, IRISA}
  \city{Rennes}
  \country{France}
}

\author{Cl\'ementine Maurice}
\email{clementine.maurice@inria.fr}
\orcid{0000-0002-8896-9494}
\affiliation{%
  \institution{Univ. Lille, CNRS, Inria}
  \city{Lille}
  \country{France}
}

\renewcommand{\shortauthors}{Fayolle et al.}

\begin{abstract}
    Microarchitectural side-channel research has grown at an exceptional pace in recent years, increasing the need for rigorous and meaningful benchmarking.
    Early attack papers typically relied on indirect proxies, such as covert-channel bandwidth or key-recovery on naive AES and RSA implementations, setting de facto standards that many subsequent works continued to replicate, sometimes by directly comparing against raw numbers from prior work.
    While these practices offer convenient points of comparison, current benchmarks may not be the most relevant to assess specific properties of new primitives. 
    Even more problematic, microarchitectural attacks are notoriously sensitive to experimental conditions: minimal changes in the target system can significantly alter outcomes and performance. 
    As a result, inadequate evaluation practices undermine reproducibility and cast doubt on the relevance of comparisons, even in top-tier venues where such issues should be identified.
    
    This paper tackles the core problem of proper benchmarking for microarchitectural side-channel attacks and examines its broader impact on research quality in the field.
    We survey 83 attack papers published in top-ranked security and architecture conferences from 2014 to 2024.
    From this corpus, we identify and define 19 recurrent benchmarking flaws that affect evaluation completeness, relevance, soundness, and reproducibility. 
    These flaws include unfair or absent comparisons, missing code or materials, and the failure to evaluate the key attack properties. 
    On average, each paper exhibits 5.72 such flaws, highlighting how widespread the issue is, even in highly selective venues.
    Based on our findings, we identify and suggest key properties that are relevant to properly evaluate new attacks.
    We also highlight trends over time and different practices between security and architecture conferences.
\end{abstract}


\maketitle

{\small
© 2026 Copyright held by the authors. This is the authors' version of the work. It is posted here for your personal use. Not for redistribution. The definitive Version of Record is published by ACM.
\\DOI: \url{https://doi.org/10.1145/3830454.3846555}}

\section{Introduction} \label{sec:introduction}

Within the last 20 years, we have witnessed a growing interest in microarchitectural side channels.
Broadly speaking, contributions in this area fall into two categories: those that focus on the \textit{software} side by uncovering and exploiting constant-time violations in real implementations \cite{benger2014ooh,pereida2016make,garcia2017constant,hassan2020deja,de2020dragonblood,de2021parasite,de2023from}, and those that focus on the \textit{hardware} side by introducing new microarchitectural primitives or mechanisms that enable information leakage. 
In this work, we concentrate on the latter category, where contributions aim to characterize or exploit microarchitectural behavior directly.
The novelty of an attack may stem from leveraging a previously unexplored component~\cite{aciiccmez2007power,aldaya2019port}, improving an existing technique~\cite{evtyushkin2018branchscope,yarom2014flush}, or porting a known attack to a new architecture~\cite{lipp2016armageddon}. 
Regardless of the specific contribution, any work must rigorously evaluate its methodology and the performance of its various features through a well-designed benchmark.
Despite the central role of evaluation in these contributions, a fundamental question remains unanswered: \textbf{How should we meaningfully evaluate and compare microarchitectural side-channel attacks?}

A first issue is that \textit{state-of-the-art benchmarks are based on historical practices that may no longer be suitable}.
They often boil down to comparing against ``classical'' attacks (\eg attacking an AES T-table implementation or ``square-and-multiply'' modular exponentiation) and trying to maximize the bit rate of a covert channel ~\cite{yao2018are, purnal2021prime, katzman2023gates}.
However, benchmarks exploiting a long-known and patched vulnerability are of limited relevance in modern implementations\footnote{\url{https://www.redhat.com/en/blog/its-all-question-time-aes-timing-attacks-openssl}{}}.
In practice, the vulnerability being exploited is not the point of the evaluation; it is merely a proxy to demonstrate that the side channel has sufficient resolution to attack non-constant-time cryptographic implementations.
Nevertheless, this practice has become almost universal in the community, with little reflection on whether such benchmark examples remain valid or meaningful.

A second issue is that \textit{it is not possible to compare ``apple-to-apple'' different side channels based on publications alone}. 
Indeed, implementations vary from one paper to another, and therefore the outcome of a specific primitive may vary dramatically across implementations. 
For example, a covert channel that integrates an error-correction mechanism and a full protocol stack~\cite{maurice2017hello} cannot be meaningfully compared to one that simply transmits raw bits with a 20\% error rate.
Moreover, experimental conditions vary across works, and, because microarchitectural side channels are notoriously sensitive to noise, results obtained in a controlled “lab’’ setup cannot be meaningfully compared to results obtained in more realistic conditions, \eg in a cloud environment.

A third issue is \textit{reproducibility}. To mitigate the second issue, researchers sometimes attempt to run state-of-the-art side channels on their experimental setup to enable meaningful comparison. However, code is not always available or working as expected. 
Artifact evaluations (AE) and the increasing requirement to release artifacts are gradually improving this situation, but dozens of previously published side channels still need to be reimplemented by authors---an important engineering 
effort that is not always well recognized.
Beyond the engineering workload, academic publications frequently omit basic engineering ``details'', which may be considered scientifically minor but are nevertheless crucial for practical reproduction.
Last, microarchitectural side channels themselves are intrinsically difficult to reproduce. 
For example, any change in a new processor generation may introduce undocumented modifications in components, causing an attack to stop working or requiring significant adaptation.

In this paper, we provide an overview of the evolution and current state of benchmarking practices in microarchitectural side-channel research from 2014 to 2024.
We studied 83 papers published in eight leading conferences in this field. 
Specifically, we selected papers from top security venues (CCS, NDSS,  Security \& Privacy, and USENIX Security Symposium) and top architecture conferences (ASPLOS, HPCA, ISCA, and MICRO).
Based on this corpus, we identify and classify the features that are commonly evaluated and emphasized, both explicitly and implicitly.
We also introduce a taxonomy that captures the key properties that should be exposed and assessed when evaluating microarchitectural side channels.
Finally, we provide a checklist to help researchers benchmark their attacks in a way that yields meaningful insights into the actual capabilities of their primitives.
In the future, this checklist could help establish a unified benchmark for the rigorous evaluation of microarchitectural side-channel research.

In a nutshell, our contributions are the following:
\begin{compactenum}
    \item We systematically study 83 papers presenting microarchitectural side channels, published in top conferences from 2014 to 2024. We highlight common practices in the evaluation of microarchitectural side channels.
    \item We highlight multiple flaws that hinder completeness, relevance, soundness, or reproducibility of microarchitectural side-channel attacks, and complicate their reproducibility.
    \item We analyze trends over time, and differences between security and architecture communities.
    \item We propose a set of recommendations to reduce improper benchmarking and mitigate its negative impact on research quality in the field.
\end{compactenum}

We share all data and scripts in~\Cref{appendix:data}.

\section{Motivation} \label{sec:Motivation}

Our aim throughout this article is not to name and shame individual works, but to highlight recurring practices that meaningfully reduce the value of empirical evidence.
The examples below are representative and motivated our work: they are in no way a judgment on the merit or quality of the cited papers.

Even within a single primitive such as \PrimeProbe, initially introduced in 2005~\cite{osvikST06}, reported numbers span a wide range because platforms, protocols, and threat models differ.
For instance, in~\cite{maurice2017hello}, a cloud-oriented resilient implementation of the covert channel reporting tens of kilobytes per second with strong error correction and no observable error in a realistic setting sits alongside more controlled evaluations that may optimize the speed at the expense of error rate (see~\cite{maurice2017hello}, Table 1).
These inaccuracies may be exacerbated when comparing new results against raw numbers of older papers that lack most recent optimizations~\cite{cui2022abusing}.
Likewise, some works~\cite{kim2022dprime} choose to compare to a single other work reporting a similar bandwidth, measured in a substantially different setting, to justify the relevance of their channel.
Without normalizing  the evaluation parameters and accounting for most recent progress, comparing these raw bandwidths is at best meaningless, and may mislead readers.

Another central source of confusion, obscuring trade-offs with other properties, is the notion of ``error rate'', which is defined differently across papers.
Several works measure errors using the Edit distance (Levenshtein distance, or Wagner-Fisher algorithm~\cite{navarro2001guided}) to capture bit flips, insertions, and deletions~\cite{xiong2020leaking,zhang2023tunnels,cui2022abusing,deng2022leaky}.
Others argue that using Hamming distance (which only captures bit flips) better reflects classical error-correcting capabilities~\cite{xiong2021leaking}.
Other works define error via accuracy or report an error rate without specifying the computation~\cite{dutta2023spy,wang2022hertzbleed}.
Since Edit and Hamming distances capture different transmission failures, ``x\% error'' conveys different meanings depending on the definition.
Unless the metric and any ECC are explicitly defined, fair comparison is impossible.

\section{Methodology} \label{sec:methodology}

In this section, we present our detailed methodology, the definition of our criteria, and the design of a systematic review pipeline for selected papers.
We describe our process for selecting and discarding papers, coding qualitative data while accounting for claim-specific context, and conducting statistical analyses of quantitative data.
We also discuss the limitations and potential biases of our work.

\subsection{Survey methodology}

To conduct a consistent meta-analysis of benchmarking practices, we first designed a structured survey and a generic codebook to guide paper annotation. 
The survey captures recurring elements of a paper's evaluation, such as the type of primitive introduced, the stated evaluation goals, the targets and platforms considered, the reported metrics, the comparisons performed, and the availability of artifacts. 
The purpose of this template is to provide a common vocabulary for analysis across papers, \emph{not to impose a rigid checklist irrespective of the contribution}.

Our methodology follows a "paper claims first, survey criteria second" approach.
For each paper, we first identified the contribution as framed by the authors and the scope of the claims supported by the evaluation. 
We then applied the flaw criteria conditionally, assessing whether the absence of a benchmark, comparison, or other relevant information weakened the evidence supporting the paper's stated claims.
This distinction is important because microarchitectural papers do not all aim to demonstrate the same kind of result. 
Some papers introduce a broadly applicable primitive and therefore require evidence about portability, robustness across settings, or comparison to related work. 
Others intentionally focus on a mechanism tied to a specific platform, component, or microarchitecture family. 
In the latter case, a broader evaluation may strengthen the paper, but it is not necessarily required to support the contribution as claimed. 
For instance, if a paper explicitly studies a feature specific to the Apple M1, evaluating the attack on a single target is not, by itself, grounds for marking the paper as flawed under the ``single platform'' criterion. 
In such cases, the criterion is treated as not flawed, unless the paper makes broader portability or generality claims.

With this approach in mind, we developed the survey and codebook iteratively. 
Our group of evaluators, composed of four academic researchers with prior experience in microarchitectural side-channel research (2-10 years of experience, with multiple publications in recognized venues), first discussed the paper selection criteria and an initial list of benchmarking issues expected in the corpus. 
All evaluators, also designated as experts, are authors of this article. 
We then performed a pilot annotation on a set of papers and refined both the survey and the codebook to better match the diversity of evaluation styles encountered in practice. 
During the full review process, one researcher performed the initial extraction for each paper, recording both structured answers and supporting evidence from the paper, and, when relevant, from its published artifact. 
We did not treat information as missing when the main paper explicitly delegated implementation details, parameters, or scripts to an accompanying artifact.

To improve robustness, a second researcher then reviewed the full annotation against the paper and checked whether they reached the same conclusion. 
Disagreements were resolved by re-reading the relevant passages and discussing the case among the coauthors until reaching a common interpretation. 
Throughout this process, we adopted a conservative resolution strategy: when a criterion did not clearly apply to the paper's claims, or when the evidence was ambiguous, we preferred leniency over strictness.
The codebook, used as a template for annotation, is available in \Cref{appendix:codebook}.

\subsection{Selection of papers to review}

Selecting representative papers is essential for a meaningful study of benchmarking trends. 
Microarchitectural side-channel attacks have become a prolific research field, especially over the last two decades.
We studied papers discussing new side channels or improvements to existing methods, focusing on the microarchitectural aspect, and not the constant-time violations in specific software implementations. 
Specifically, we discarded any article whose core contribution is to attack a specific cryptographic implementation. 
Similarly, we excluded articles focusing exclusively on speculative execution attacks, such as Spectre, Meltdown, and MDS-based attacks, as they reuse existing side channels for secret exfiltration, but their core contribution is not about the side channel itself, but rather traces left by transient execution.

Microarchitectural security is mainly studied by two closely related research communities: security-focused and architecture-focused.
Similar to the methodology from~\cite{KouweHABG19}, we restricted our selection to the main conferences in each field based on their CORE2026 rankings.
Although CORE ranking may not be the best metric, it is well-established as a metric influencing a paper's visibility and impact.
Similarly, these venues have selective reviewing and \emph{should} filter out contributions with obviously flawed evaluations.
Ultimately, we focused on the following venues: ASPLOS, HPCA, ISCA, and MICRO for architecture conferences, and CCS, NDSS, Security \& Privacy, and USENIX Security Symposium for security conferences.

We focused on articles published within the last decade, from 2014 to 2024. 
We chose 2014 as the starting date because it coincided with the publication of a keystone in the field: the \FlushReload attack~\cite{yarom2014flush}, which arguably renewed interest in the field.
We ended our search in 2024 to cover the ten-year period. 

We systematically scraped the proceedings of all the selected conferences, manually excluding any paper that did not evaluate a side or covert channel, yielding a corpus of 114 papers further refined to 83 to match our scope.
Four experts performed this task independently and discussed any uncertainties about the scope of the papers until they reached an agreement.
An extensive list of the papers selected for this study is available in \cref{appendix:papers}, sorted by year, venue, and title.

\subsection{Limitations}

We consider multiple limitations to our methodology.
First, we ensured that experts did not evaluate their own papers. However, as experts in the field, we may still be biased due to our knowledge and experience.
For example, we may be subject to opt-in or self-reporting biases.
To limit these effects, we cross-checked all our annotations, with a different expert proofreading and checking for mistakes or inaccuracies. 
If they could not reach a consensus after discussing it, all four experts gathered to decide on the final answer.

Despite these precautions, we do not claim our list of flaws to be complete or sound, as compiling such a list is not the main goal of our paper.
Our selection of top conferences might introduce some bias in our results, as it excludes papers from other well-known venues. 
However, this selection was necessary to keep our corpus at a manageable size, and still covers dozens of papers from the security and architecture communities.
Moreover, we note that our corpus is composed of more papers from security conferences than architecture conferences, and that more recent papers are represented.
While this imbalance may be the result of differing interests in side-channel attacks, it may introduce some bias in our results.
We hope our sample represents the most impactful and visible papers in both fields.
Similarly, our scope for included attacks may lead to some flaws being omitted, as we exclude transient execution attacks and papers whose novelty lies in exploiting constant-time violations in implementations.
Benchmarking practices in these papers, particularly for speculative attacks, tend to be different from side-channel attacks and evaluating them deserves its own paper.

Lastly, we merged attacks with similar characteristics and targets introduced within the same paper in a single occurrence.
This was necessary as some papers did not clearly distinguish the number of attacks they used for evaluation. 
This has a minimal impact on the total number of attacks evaluated. 

\section{Taxonomy of flaws} \label{sec:flaws}

To build our taxonomy, we first define criteria expected from a good evaluation. 
From there, we can identify the aspects in which the various benchmarks differ from a ``perfect'' evaluation.
Following the reasoning of van der Kouwe \textit{et al.}~\cite{KouweHABG19},  we expect a good  evaluation to meet the following properties.

First, it should be \textbf{\hypertarget{complete}{complete}}, \ie it should properly assess all key properties of the novel attack primitive, and highlight their limitations.
We emphasize that we do not consider completeness as measuring everything, but \emph{stating and measuring the right things}.
Second, it should be \textbf{\hypertarget{relevant}{relevant}}, \ie conveying meaningful information without requiring inference from the reader. 
Third, it should be \textbf{\hypertarget{sound}{sound}}, meaning that the results must accurately measure the intended properties rather than reflecting unintended side effects or very controlled settings only. 
Last, the evaluation should be as \textbf{\hypertarget{reproducible}{reproducible}} as possible.
Readers should have enough elements to reproduce the experiment and the benchmarks on their own, at least on the same device. 

\begin{table}[t]
    \scriptsize
    \caption{Flaws and their impact on evaluation properties.}
    \label{tab:flaws}
    \centering

    \begin{threeparttable}[h]
        \rowcolors{2}{gray!12}{white}
        \begin{tabular}{p{4.5cm}cccc}
            \toprule
            \textbf{Flaw} & \textbf{\hyperlink{complete}{Com.}} & \textbf{\hyperlink{relevant}{Rel.}} & \textbf{\hyperlink{sound}{S.}} & \textbf{\hyperlink{reproducible}{Rep.}} \\
            \midrule
            \codebookref{A1}\label{flaw:A1} No error rate                                                    & \CIRCLE     & \CIRCLE &             &             \\
            \codebookref{A2}\label{flaw:A2} No spatial resolution                                            & \CIRCLE &             &             &             \\
            \codebookref{A3}\label{flaw:A3} No temporal resolution                                           & \CIRCLE &             &             &             \\
            \codebookref{A4}\label{flaw:A4} No evaluation of noise resilience                                & \CIRCLE & \CIRCLE     &             &             \\
            \codebookref{A5}\label{flaw:A5} No evaluation against mitigations the paper claims to bypass     & \CIRCLE & \CIRCLE &             &             \\
            \codebookref{A6}\label{flaw:A6} Evaluated on a single platform                                   & \CIRCLE &             &             & \CIRCLE     \\
            \codebookref{A7}\label{flaw:A7} Evaluated on a single configuration                              & \CIRCLE &  &             &             \\
            \codebookref{B1}\label{flaw:B1} Unfair comparisons                                               &             & \CIRCLE     &  \CIRCLE     &             \\
            \codebookref{B2}\label{flaw:B2} No comparison                                                    &             & \CIRCLE &             &             \\
            \codebookref{B3}\label{flaw:B3} Evaluation designed to highlight performances                    &             & \CIRCLE & \CIRCLE     &             \\
            \codebookref{B4}\label{flaw:B4} Benchmark only on highly controlled/simplified environments      &             & \CIRCLE & \CIRCLE &             \\
            \codebookref{C1}\label{flaw:C1} Missing target/target version                                    &             &             &             & \CIRCLE \\
            \codebookref{C2}\label{flaw:C2} Missing CPU model                                                & \CIRCLE     &  &  \CIRCLE  & \CIRCLE \\
            \codebookref{C3}\label{flaw:C3} No code / material / documentation to reproduce                  &             &             &             & \CIRCLE     \\
            \codebookref{C4}\label{flaw:C4} No number of samples for side-channel attacks                    &             & \CIRCLE &             &             \\
            \codebookref{C5}\label{flaw:C5} Missing details on classification of side-channel traces         &             &             &             & \CIRCLE \\
            \codebookref{C6}\label{flaw:C6} Missing protocols for covert channels                            & \CIRCLE &             &             & \CIRCLE \\
            \codebookref{C7}\label{flaw:C7} Missing information on error rate computation / error correction &             &             & \CIRCLE & \CIRCLE \\
            \codebookref{C8}\label{flaw:C8} Missing prerequisites                                            & \CIRCLE &      &             & \CIRCLE \\
            \bottomrule
        \end{tabular}
    \begin{tablenotes}[flushleft]
    \footnotesize
    \item \textbf{\hyperlink{complete}{Com.}}: Completeness \quad \textbf{\hyperlink{relevant}{Rel.}}: Relevance \quad \textbf{\hyperlink{sound}{S.}}: Soundness
    \item \textbf{\hyperlink{reproducible}{Rep.}}: Reproducibility
    \quad \CIRCLE~denotes an impact on the corresponding property.
    \end{tablenotes}
    \end{threeparttable}
\end{table}

We adapted and complemented a list of flaws from previous work~\cite{KouweHABG19} to fit the specificity of our field. 
Based on our experience and this large-scale and systematic survey (\Cref{sec:results}), we define 19 flaws that may impact properties.
We summarize the most impacted properties for each flaw in~\Cref{tab:flaws}.

Nonetheless, it is important to note that not all papers are affected to the same extent.
We used a range of labels to describe the impact of a flaw on an evaluation. 
We started with a generic definition of these labels and refined it based on each flaw when necessary (see~\Cref{appendix:codebook}).
Specifically, we used the following labels:
\begin{itemize}
    \item \okKeyword: the requirement is met, and the evaluation is not affected by the flaw.
    \item \emph{Underspecified -} \unclearKeyword : the idea is discussed, but it is scattered or implicit. An informed reader could infer it, but it is not stated plainly.
    \item \emph{Underspecified -} \incompleteKeyword: the information is partially provided, but elements are missing, or only provided for some evaluations.
    \item \flawedKeyword: clearly wrong, misleading, or does not meet the requirement.
    \item \naKeyword: Not Applicable to this paper (\eg covert-channel protocol details in a pure side-channel paper).
\end{itemize}

In the following subsection, we describe the considered flaws and their impact, grouped into three categories corresponding to the following subsections.

\subsection*{A. Benchmark omissions}

The first category of flaws denotes any evaluation practices that may hide or skip (either intentionally or unintentionally) any experimental dimensions that affect a relevant property of the attack primitive.
Microarchitectural channels live at the intersection of hardware, firmware/microcode, OS and application behavior.
Any parameter across layers may interact with multiple aspects of the channel: sampling rate, thresholds, affinity, etc.
However, it is impractical to exhaustively cover this space.
The goal is to provide a principled evaluation that \textit{(i)} states what is being measured, \textit{(ii)} quantifies key trade-offs, and \textit{(iii)} enables future normalization.

Omitting elements affecting properties such as reachable sampling frequency, noise resilience, or error rate impacts the evaluation's quality and does not allow for a proper comparison, making primitives appear universally strong.
We refer to these as \emph{benchmarking omissions}.
Ultimately, benchmarking omissions obscure trade-offs that are essential for scientific comparisons.

When assessing channel reliability, reporting the error rate is key. 
The unit (bit/byte) and an appropriate number of trials must be explicitly stated. 
Vague statements such as ``zero-error'' are insufficient in that regard. 
Flaw \textbf{\flawref{A1}}\label{code_to_flaw:A1} covers such issues. 
Failing to report error rates allows high throughput to mask errors.
This flaw mostly impacts \emph{completeness}, but \emph{relevance} may also be weakened since the reader cannot properly assess the attack's practicality.

Spatial and temporal resolutions are two keystones for evaluating attack potency.
The former describes the smallest distinguishable unit and the mapping from observation to unit (may it be cache line, set, bank, page, port, etc.).
The latter covers the time-per-sample and maximal sustainable sampling rate. 
Knowing the theoretical and practical limitations of such dimensions improves the reader's understanding of the attack and its potency, offering insights on how it may compare to other attacks.
A lack of proper characterization of these properties is described by flaws \textbf{\flawref{A2}}\label{code_to_flaw:A2} and \textbf{\flawref{A3}}\label{code_to_flaw:A3}, respectively.
In both cases, the completeness of the evaluation is impacted: key properties that should be explicit are missing, hindering comparison with other works. 

While theoretical resolution is important, it is not sufficient. 
Quantifying a primitive's \emph{noise resilience} under realistic events that may impact its performance provides an insight on the resilience of the attack.
Indeed, a primitive working flawlessly on a target, with skyrocketing error rates as soon as another process is running may not be comparable to a more resilient attack.
A lack of noise resilience evaluation (flaw \textbf{\flawref{A4}}\label{code_to_flaw:A4}) impacts both evaluation \emph{completeness} and, to a greater degree, \emph{relevance}.
Without a performance degradation curve under system stress, we may not understand that the attack works only in  quiescent setups. 

Flaw \textbf{\flawref{A5}}\label{code_to_flaw:A5} is an instance of an unsubstantiated claim.
Any claim unsupported by evidence should not be included in a scientific article.
It \emph{affects} both the completeness of the evaluation and its \emph{relevance}, as the claim should be disregarded.

Portability is especially challenging in microarchitectural attacks.
While seminal papers were often content with an attack working on a specific platform, it has been widely acknowledged that a generic attack should work on multiple platforms.
Even when full portability is not the ultimate goal, studying which platform may be affected by a new primitive is key to conceptualizing the attack surface.
In this regard, we expect key behaviors causing the attack to be replicated on different architectures, and a description of necessary adaptations.
Restricting the entire evaluation to a single platform (\textbf{\flawref{A6}}\label{code_to_flaw:A6}) impacts the evaluation's \emph{reproducibility} significantly, as well as its \emph{completeness}. 

Finally, studying various primitive configurations is required to test the attack's sensitivity to such parameters.
We consider a change of configuration as a sweep of at least one impactful (\ie affecting observable outcomes) factor during the evaluation. 
This includes changes in the noise level, different thresholds, or window sizes, etc.
This is only required to find the best settings for the attack, and we are not expecting all further evaluation based on the same primitive to be done in multiple configurations.
However, if a paper introduces different attack primitives (based on different behaviors, or different components), each should be evaluated in multiple configurations.
An evaluation performed on a single configuration (\textbf{\flawref{A7}}\label{code_to_flaw:A7}) would focus on a single aspect, lacking \emph{completeness}.

\subsection*{B. Improper comparisons}

An evaluation compares \emph{properties}, under specific \emph{conditions}, using some \emph{metrics}.
A proper comparison should be both qualitative and quantitative.
Improper comparisons may arise when results are compared across incompatible protocols, hardware, threat model, or any measurement semantics without normalization.
This issue is especially relevant in microarchitectural security: any small change in the evaluation target or setup may drastically alter observations.
Combined with setup heterogeneity, it becomes particularly difficult to compare without \textit{(i)} reproducing the attack we are comparing to, or \textit{(ii)} explicitly normalizing and qualifying the differences.

In a sense, an unfair comparison (\textbf{\flawref{B1}}) may be worse than none at all (\textbf{\flawref{B2}}), as it creates a false sense of superiority, when it is in fact comparing different elements.
On the one hand (\textbf{\flawref{B1}}\label{code_to_flaw:B1}), comparing raw numbers without accounting for a different setup, or generic improvement brought by recent works could mislead a reader into thinking a contribution is superior, while the comparison is not meaningful.
This would therefore impact the \emph{relevance} of the evaluation.
In some cases, the \emph{soundness} may be impacted, as we are no longer comparing the primitives' properties, but rather the incidental platforms (\eg different cache size, higher clock, etc.).
On the other hand (\textbf{\flawref{B2}}\label{code_to_flaw:B2}), absence of comparison prevents readers from positioning the contribution in the literature, impacting the \emph{relevance} of the evaluation.
We acknowledge that first-of-its-kind attacks may not be comparable, and were labeled \emph{NA}.

Evaluations that use ``easy'' payloads (\eg strong repeating patterns) or extrapolate from partial attack stages to non-trivial end-to-end scenarios are designed to highlight performance (\textbf{\flawref{B3}}\label{code_to_flaw:B3}).
In the former case, repetitive payloads may trigger additional microarchitectural effects; in the latter, they may lead to incorrect conclusions due to oversimplified reasoning.
This significantly impacts \emph{soundness} and may also undermine practical \emph{relevance}.

Similarly, benchmarking only in highly controlled or simplified setups (\textbf{\flawref{B4}}\label{code_to_flaw:B4}) also impacts the evaluation's \emph{soundness} and \emph{relevance}.
Evaluating an attack only in idealized systems, or with prerequisites inconsistent with the threat model, may measure properties related to the setup itself, not the attack.

\subsection*{C. Missing information}

This category of flaws prevents readers from reconstructing the experiments or understanding precisely what is measured.
As stated before, minor setup differences can have cascading effects, breaking the chain from phenomenon to measurement.
Hence, this category has a significant impact on \emph{reproducibility} and, by extension, on future comparability.

Omitting OS, kernel, or targeted software version (\textbf{\flawref{C1}}\label{code_to_flaw:C1}) directly impacts \emph{reproducibility}.
Indeed, any library update could introduce subtle changes that impact the attack. 
Two systems with the ``same browser'' may implement different timer restrictions depending on the version~\cite{rokicki2021sok}.

Similarly, two machines with the ``same CPU/GPU family'' may behave widely differently due to microarchitectural changes.
While \textbf{\flawref{C1}} is about the versions in the evaluation, neglecting to specify the CPU/GPU model(s) (\textbf{\flawref{C2}}\label{code_to_flaw:C2}) directly relates to the main properties defining the attack.
In addition to the impact on \emph{soundness} and \emph{reproducibility}, it has a major impact on \emph{completeness}.

Making attack code available (\textbf{\flawref{C3}}\label{code_to_flaw:C3}) could be considered the backbone of \emph{reproducibility}.
However, merely making an artifact available is not sufficient: attack primitives rely on implementation details, and numerous elements may need to be modified and adapted to make it work.  
A proper documentation should help an external reader to not only reproduce the experiment under the same conditions, but also to get an idea of what needs to be adapted to make it work on \textit{similar} targets.
Beyond reproducibility, good artifacts enable apple-to-apple comparisons by letting future works re-run baselines under the same conditions, leading to an improvement regarding \textbf{\flawref{B1}}.

When evaluating a side channel to demonstrate an attack's potency, it is important to document the number of samples required.
This number could be reported for recovering an element (\eg how many samples are required to recover a bit of the RSA key) or for the end-to-end attack.
Most attacks do not run reliably with a single measurement, so the number of samples gives an indication of the attack's signal-to-noise ratio.
A side-channel evaluation not reporting sample numbers (\textbf{\flawref{C4}}\label{code_to_flaw:C4}) lacks some \emph{relevance} in this regard.

Many evaluations aim at classifying instances, such as fingerprinting (website, application, etc.) or side-channel trace denoising.
Classification details (model used, features, training set, etc.) are crucial to reproduce the experiments.
While the primitive would still perform in the same way, omitting these elements (\textbf{\flawref{C5}}\label{code_to_flaw:C5}) has an impact on \emph{reproducibility}.

For covert channels, the communication protocol must be well-defined and explicitly described.
Missing initialization/synchronization (\textbf{\flawref{C6}}\label{code_to_flaw:C6}), whether the channel is unilateral or bilateral, framing, clocking, and initial shared knowledge have an impact on the \emph{completeness} and \emph{reproducibility} of the evaluation.
In addition, a proper description of the protocol helps to characterize the attack model, helping comparison to other works. 

Reporting the error rate (see flaw \textbf{\flawref{A1}}) is important, but a plain error rate is not enough as various metrics can be used to compute it.
Error rates can also be computed from channel accuracy, or computed after applying various error corrections.
It is thus important to describe how the error rate was obtained.
Missing this (\textbf{\flawref{C7}}\label{code_to_flaw:C7}) would impact both the \emph{soundness} and \emph{reproducibility} of the evaluation.

Finally, some papers may introduce multiple primitives, or variants with different prerequisites (privileges, co-location, etc.).
If these changes are not stated carefully (\textbf{\flawref{C8}}\label{code_to_flaw:C8}), the threat model and evaluation prerequisites may become unclear, impacting both \emph{completeness} and \emph{reproducibility}.

\section{Survey results} \label{sec:results}

\begin{figure}[t]
    \centering
    \includegraphics[width=\columnwidth]{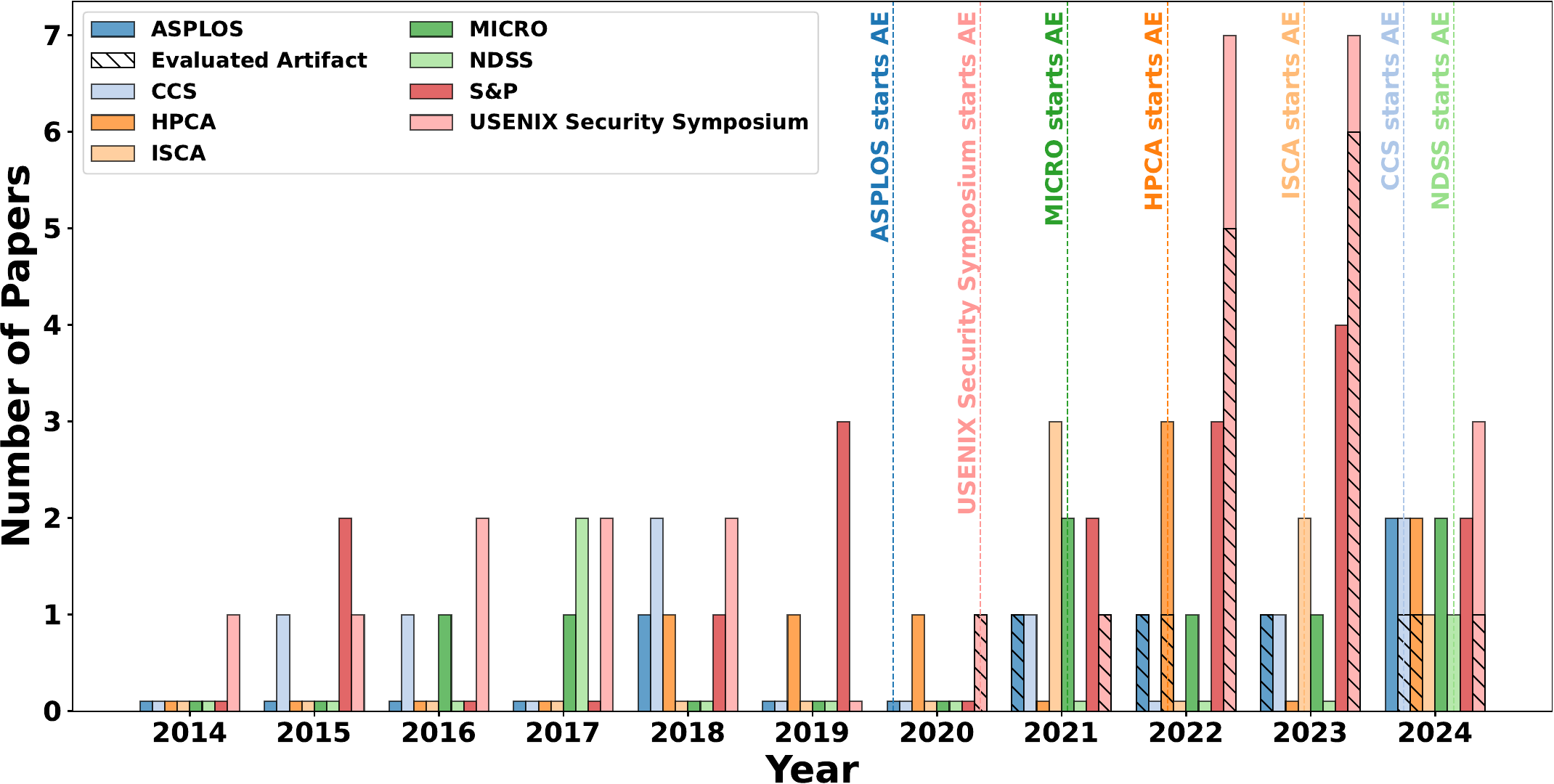}
    \caption{Number of papers and evaluated artifacts per conference per year, with the date that AE began, if applicable.}
    \label{fig:number_of_papers_and_evaluated_artifact}
\end{figure}

\begin{figure}[t]
    \centering
    \includegraphics[width=\columnwidth]{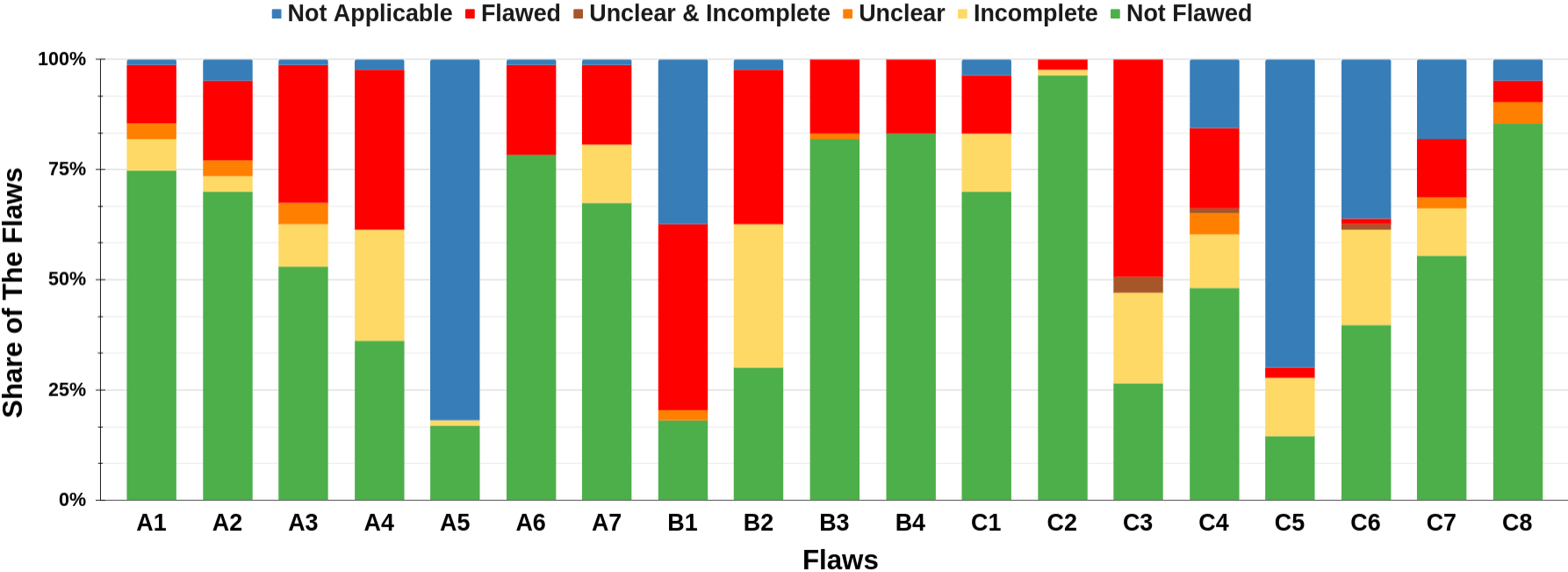}
    \caption{Prevalence of the 19 flaws in the 83 surveyed papers.}
    \label{fig:share_of_flawed}
\end{figure}

In this section, we present the results of our survey on 83 papers published between 2014 and 2024 in top security and architecture conferences, as shown in \cref{fig:number_of_papers_and_evaluated_artifact}.
We analyze the prevalence of the 19 flaws in our taxonomy across the surveyed papers, as illustrated in \cref{fig:share_of_flawed}.
To simplify the reporting of results, we use \emph{partial flaws} to  refer to \emph{Underspecified - Incomplete}, \emph{Underspecified - Unclear}, or \emph{Underspecified - Unclear \& Incomplete} flaws, indicating that the information was either partially provided, scattered, or implicit.

\begin{figure}[t]
    \centering
    \includegraphics[width=\columnwidth]{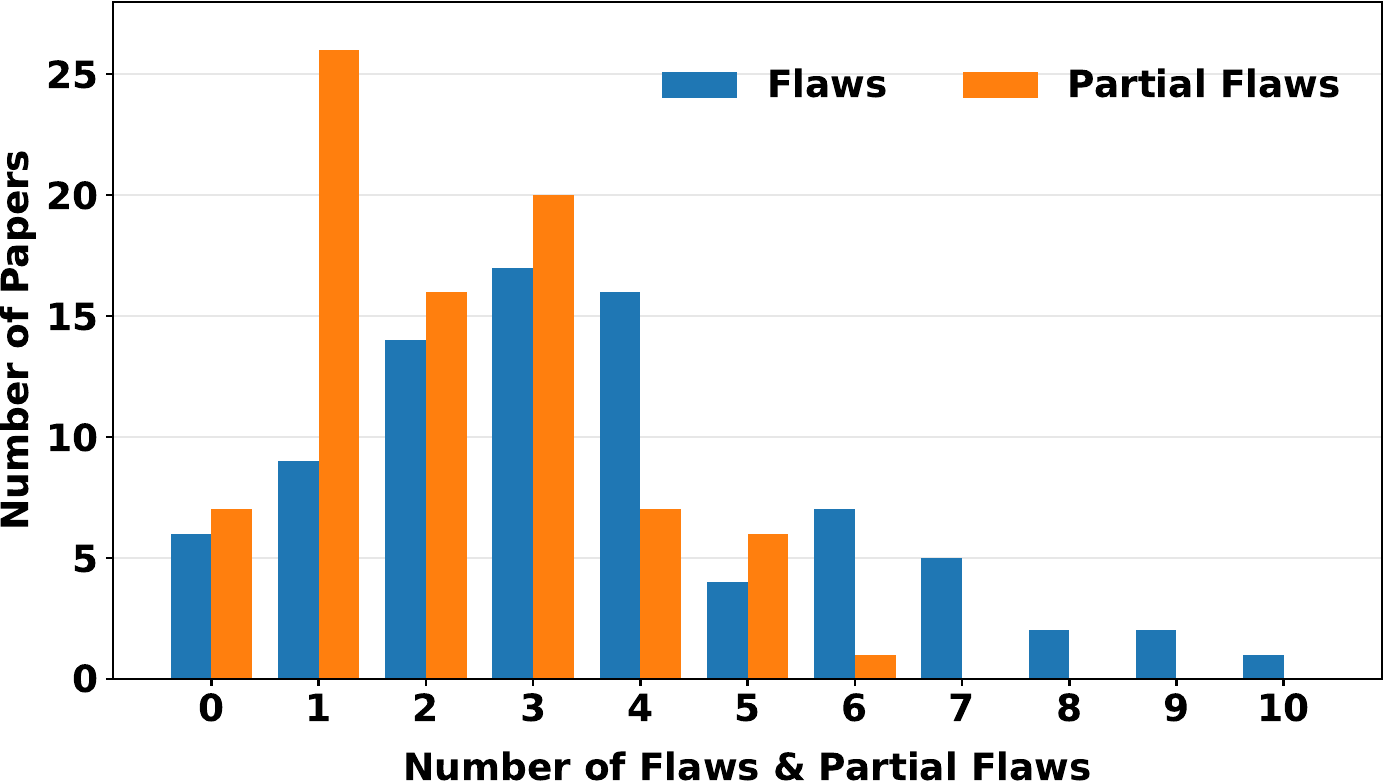}
    \caption{Distribution of the number of flaws per paper.}
    \label{fig:share_of_the_flaws}
\end{figure}

In our corpus, we found that 77 papers (92.8\%) suffer from at least one full flaw and 76 papers (91.6\%) suffer from at least one partial flaw, with a median of 3 full flaws and 2 partial flaws per paper.
The distribution of full and partial flaws per paper is shown in \cref{fig:share_of_the_flaws}.
On average, a paper suffers from 5.72 full or partial flaws.
This highlights the pervasiveness of benchmarking issues in microarchitectural research.
We distinguish the flaws based on their impact within each category (benchmark omissions, improper comparisons, and missing information), as described in \cref{tab:flaws}.

After summarizing trends from \cref{tab:summary_flaws}, we present low-\-pre\-va\-lence flaws, then focus on the most common ones, organized by the three categories of \Cref{sec:flaws}.

\subsection{General observations}

\paragraph{Covert-channels vs. side-channels distribution} 
As shown in \cref{fig:number_of_papers_and_evaluated_artifact}, our corpus contains a diverse range of papers from various conferences and years.
These papers cover different types of microarchitectural side-channel and covert-channel attacks, as well as other related contributions.
Indeed, the 83 papers describe a total of 218 microarchitectural side-channel and covert-channel attacks.
Among those, 147 (67.4\%) are side-channel attacks, while 71 (32.6\%) are covert-channel attacks.
Additionally, we identified a few other types of contributions, such as improving building blocks of widespread attacks (minimal eviction set finding, performance degradation improvement), reverse engineering efforts, Rowhammer attacks, or specific microbenchmarks.

\begin{figure}[t]
    \centering
    \includegraphics[width=\columnwidth]{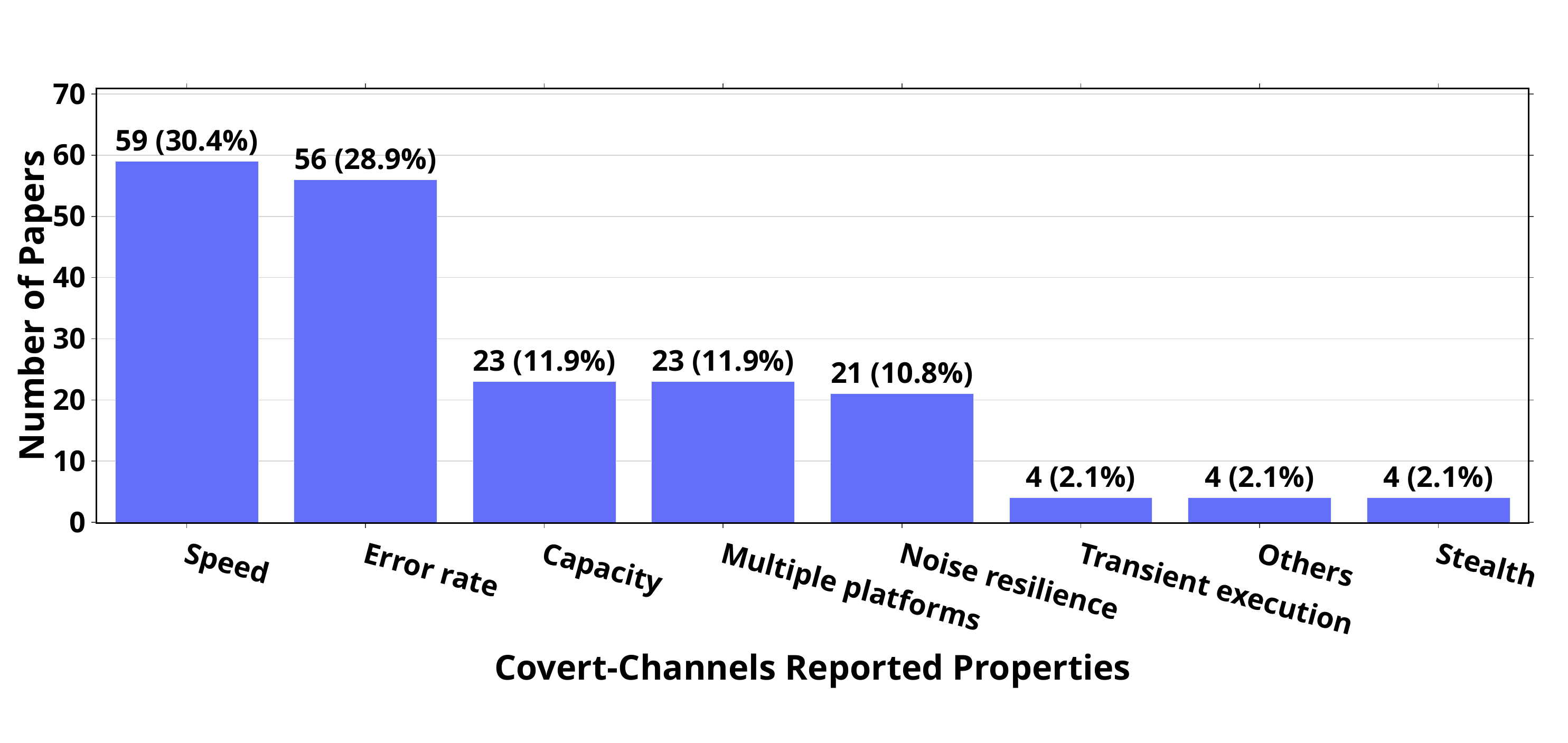}
    \caption{Distribution of reported properties in covert-channel evaluations. Percentages are normalized over all instances (\eg "Speed" represents 30.4\% of all properties).}
    \label{fig:propotion_of_covert_channel_s_assessment_type}
\end{figure}

\paragraph{Assessment of covert-channels}
For covert-channel attacks, we analyzed the most commonly reported properties of the channel, as depicted in \cref{fig:propotion_of_covert_channel_s_assessment_type}.
We found that 59 papers (71.1\%) evaluate the raw speed of their covert channels, and 56 (67.5\%) assess the error rate.
Notably, 18 papers (21.7\%) evaluate only speed and error rate.
30 papers (36.1\%) evaluate either speed or error rate in conjunction with other metrics, such as noise resilience or multi-platform performance.

\begin{figure}[t]
    \centering
    \includegraphics[width=\columnwidth]{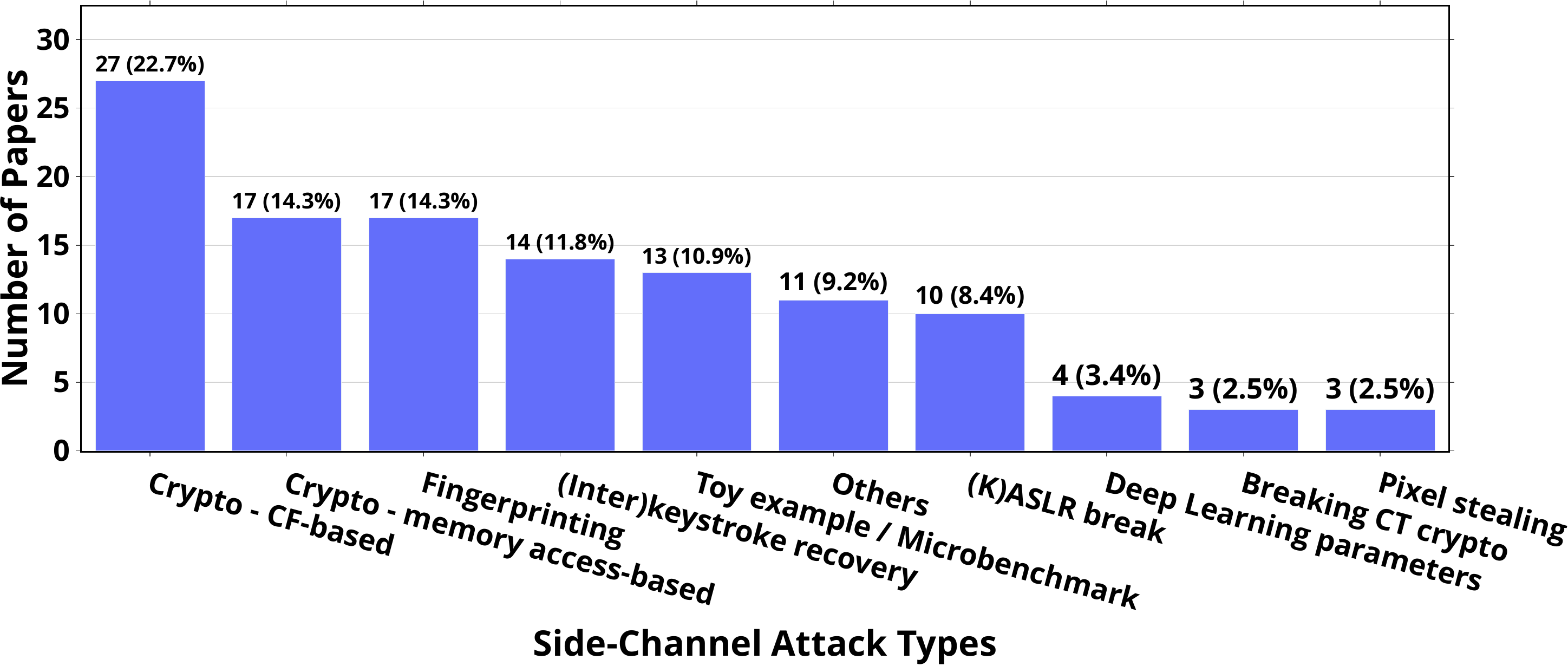}
    \caption{Distribution of side-channel attack types. Percentages show proportions across all attacks and duplicates per paper counted once (\eg "CF" represents 22.7\% of all attacks).}
    \label{fig:propotion_of_side_channel_attack_types}
\end{figure}

\paragraph{Types of side-channel attacks}
Regarding side-channel attacks, we examined the distribution of attack types in our corpus, as illustrated in \cref{fig:propotion_of_side_channel_attack_types}.
In general, we found that 27 papers (32.5\%) include at least one control-flow (CF) cryptographic attack, while 17 papers (20.5\%) feature at least one fingerprinting attack, and 17 papers (20.5\%) present at least one memory access-based cryptographic attack (\eg against AES T-table).

\paragraph{Most commonly targeted victims}
We further analyzed the most commonly targeted victims in each of these three categories of side-channel attacks in our corpus.
The results are similar to those of \cite{DBLP:journals/iacr/GeYCH16} and Lou \textit{et al.} \cite{DBLP:journals/csur/LouZJZ21} and can be found in \cref{appendix:usual_targets}.

\paragraph{Types of attacked platforms}
We also analyzed the hardware platforms used in the evaluations.
Unsurprisingly, most of the evaluated attacks were on CPUs (73 papers, representing 88\% of the corpus). 
In particular, a large majority (65 papers, 78.3\%) test at least one Intel CPU.
In contrast, we observed that 9 papers (10.8\%) evaluate their attacks on discrete GPUs, and 3 papers (3.6\%) focus on integrated GPUs.
Lastly, 9 papers (10.8\%) include evaluations on Trusted Execution Environments (TEEs).

\begin{table}[t]
\centering
\caption{Prevalence of benchmarking flaws across papers.}
\label{tab:summary_flaws}

\small
\renewcommand{\arraystretch}{0.82}
\begin{threeparttable}
\begin{tabular}{
    c
    @{\hspace{3mm}}S[table-format=3.0]
    @{\hspace{3mm}}S[table-format=3.0]
    @{\hspace{3mm}}S[table-format=2.1]
    @{\hspace{3mm}}S[table-format=3.0]
    @{\hspace{3mm}}S[table-format=2.1]
}
\toprule
Flaw &
Applicability &
\multicolumn{2}{c}{Flawed} &
\multicolumn{2}{c}{Underspecified} \\
\cmidrule(lr){3-4}\cmidrule(lr){5-6}
& \multicolumn{1}{c}{count} & \multicolumn{1}{c}{count} & \multicolumn{1}{c}{\%} & \multicolumn{1}{c}{count} & \multicolumn{1}{c}{\%} \\
\midrule

\rowseverity{13.4}
\textbf{A1} & {82} & {11} & {13.4} & {9} & {11.0} \\

\rowseverity{19.0}
\textbf{A2} & {79} & {15} & {19.0} & {6} & {7.6} \\

\rowseverity{31.7}
\textbf{A3} & {82} & {26} & {31.7} & {12} & {14.6} \\

\rowseverity{37.0}
\textbf{A4} & {81} & {30} & {37.0} & {21} & {25.9} \\

\rowseverity{0.0}
\textbf{A5} & {15} & {0} & {0} & {1} & {6.7} \\

\rowseverity{20.7}
\textbf{A6} & {82} & {17} & {20.7} & {0} & {0.0} \\

\rowseverity{18.3}
\textbf{A7} & {82} & {15} & {18.3} & {11} & {13.4} \\

\midrule
\rowseverity{67.3}
\textbf{B1} & {52} & {35} & {67.3} & {2} & {3.8} \\

\rowseverity{35.8}
\textbf{B2} & {81} & {29} & {35.8} & {27} & {33.3} \\

\rowseverity{16.9}
\textbf{B3} & {83} & {14} & {16.9} & {1} & {1.2} \\

\rowseverity{16.9}
\textbf{B4} & {83} & {14} & {16.9} & {0} & {0.0} \\

\midrule
\rowseverity{13.8}
\textbf{C1} & {80} & {11} & {13.8} & {11} & {13.8} \\

\rowseverity{2.4}
\textbf{C2} & {83} & {2} & {2.4} & {1} & {1.2} \\

\rowseverity{49.4}
\textbf{C3} & {83} & {41} & {49.4} & {20} & {24.1} \\

\rowseverity{21.4}
\textbf{C4} & {70} & {15} & {21.4} & {15} & {21.4} \\

\rowseverity{8.0}
\textbf{C5} & {25} & {2} & {8.0} & {11} & {44.0} \\

\rowseverity{1.9}
\textbf{C6} & {53} & {1} & {1.9} & {19} & {35.8} \\

\rowseverity{16.2}
\textbf{C7} & {68} & {11} & {16.2} & {11} & {16.2} \\

\rowseverity{5.1}
\textbf{C8} & {79} & {4} & {5.1} & {4} & {5.1} \\
\bottomrule
\end{tabular}
\begin{tablenotes}
\footnotesize
\item Prevalence :
\textcolor{black}{\colorbox{sev1}{\phantom{xx}} \textbf{rare} ($< 10\%$)},
\textcolor{black}{\colorbox{sev2}{\phantom{xx}} \textbf{occasional} (10\%-30\%)},
\item \quad \quad \quad \quad \quad \textcolor{black}{\colorbox{sev3}{\phantom{xx}} \textbf{moderate} (30\%-50\%)},
\textcolor{black}{\colorbox{sev4}{\phantom{xx}} \textbf{frequent} ($> 50\%$)}.
\end{tablenotes}
\end{threeparttable}
\end{table}

\subsection{Low-prevalence flaws}

While these flaws were hardly present in our corpus, they still undermine the overall quality of the evaluation.
Full flaws are rare, but partial flaws are more frequent and collectively affect the trustworthiness and clarity of reported results.

\paragraph{A5 No evaluation against claimed mitigations they bypass}
Among the 15 papers addressing this flaw, only 1 (6.7\%) claims to bypass existing mitigations, yet evaluates only a subset of them, leaving the remaining claims unverified.

\paragraph{C2 Missing CPU model}
In the entire corpus, three papers (3.6\%) omit or partially report the CPU model, complicating reproducibility and generalization across platforms.

\paragraph{C5 Missing classification details}
Among the 25 papers using classification, only 2 (8\%) provide no details at all about their classification's model.
Although 1 of these works describes its training and validation sets, and specifies what it evaluates, it omits essential information about the classifier itself, such as the model type, selected features, windowing strategy, or hyperparameters—making the experiments difficult to reproduce.
11 other papers (44\%) report this information only partially.
Overall, machine-learning-based attacks frequently lack crucial classifier details, hindering full reproducibility.

\paragraph{C6 Missing covert-channel protocol}
Among the 53 papers addressing this flaw, only 1 (1.9\%) provides no protocol details at all.
However, 19 papers (35.8\%) report these details only partially, typically because they describe some aspects of the protocol while omitting others, or document the protocol for one covert channel but not for another.
In several cases, the description is incomplete, for instance lacking information on synchronization, calibration, or transmission steps, which further hinders reproducibility.

\paragraph{C8 Missing prerequisites}
Among the 79 concerned papers, 8 (10.1\%) fully or partially fail to indicate changes in prerequisites between attacks, typically shifts in shared-memory assumptions or platform requirements.
In these cases, missing hardware information or unclear assumptions make it difficult to determine the exact conditions under which each attack operates.

\subsection{A. Benchmark omissions}

\paragraph{A1 No error rate}
13.4\% of papers fully miss the error rate evaluation, meaning they did not provide any information on the error rate itself, the accuracy, the true/false positive rates, or any related metric that would give insight into the reliability of their results.

Among the 11\% of papers that partially report this information, most acknowledge the presence of errors without providing any clear quantitative evaluation.
Information might be hidden in other global metrics like the F1-score or the capacity.
Sometimes, only a few of the benchmarked attacks of the paper report their error rate.

Other works took care to describe the evaluation of the bit error rate in detail~\cite{schaik2018malicious} or use various kinds of error rate metrics (\eg accuracy, precision, and F-score) for each of their attacks~\cite{gerlach2023security}.

\paragraph{A2 No spatial resolution}
19\% of papers do not attempt to determine true spatial resolution.
Some of them claim to distinguish between states of instructions, such as in a loop of various instructions or subtle variations in frequency due to power consumption, but their measurements are not granular or reproducible enough to be considered spatial resolutions.
1 paper clearly applies a prior attack with a defined spatial resolution to other cryptosystems and attacks, but the spatial resolution is not recalled, making it flawed.

Only 7.6\% of papers report incomplete or unclear resolution information. 
Most provide descriptions (\eg uop cache usage, distinguishing seq0/seq1, “zebra vs. tiger”-type functions) suggesting spatial resolution \textit{exists}, but without sufficient detail for proper evaluation.
Some also discuss improving spatial resolution without providing numerical values or a concrete evaluation.

Papers like \cite{purnal2021prime, briongos2020reload, gruss2015cache, guo2023uncore} indicate clear spatial resolution granularity or metrics such as the cache set size, a whole cache line or the evolution of frequency of cores.

\paragraph{A3 No temporal resolution}
31.7\% of papers omit temporal resolution or limit themselves to discussing cycles or bandwidth.
However, this cannot be considered temporal resolution due to its dependence on the setup of attacks.
Additionally, 1 paper applies a prior attack to other cryptosystems or attacks with the same properties without recalling its resolution.

14.6\% of papers give incomplete or unclear temporal resolution, in most cases because they do not give any concrete measurements or thresholds.
For example, although they refer to estimating thresholds (\eg distinguishing ``1'' vs. ``0''), no numerical values or figures are provided.
Plus, some papers summon magic numbers (\eg transmission intervals values) without evaluation or explanation.

Good examples of temporal resolution reporting appear in different forms.
In~\cite{zhang2023tunnels}, the authors search for the delay that maximizes error-free exfiltration speed.
In~\cite{bulck2017telling}, they dedicate a full subsection to the microbenchmark of their Inter-Processor Interrupt (IPI) latency, used as a temporal resolution.
In~\cite{khatamifard2019powert}, they define the theoretical transmission rate and use it for the evaluation of their attack, giving both the best theoretical bound and practical values.
Another approach consists in assessing the evolution of the error rate considering various transmission rates, giving a good idea of the temporal resolution~\cite{xiong2020leaking}.

Altogether, among 79 papers concerned by flaws A2 and A3 we found that 9 papers (11.4\%) miss both spatial and temporal resolutions fully or partially.
These issues affect the completeness of the paper and tend to have cascading effects, rippling into other evaluation weaknesses.

\paragraph{A4 No evaluation of noise resilience}
This flaw fully affects 37\% of papers, and partially affects 25.9\% of papers.
Among the partially affected papers, several discuss noise conceptually or describe how it could be mitigated, but do not actually measure it using clear metrics (\eg error rates, accuracy).
Some evaluate noise resilience only for a subset of their attacks, leaving others unevaluated.
Others state that their attack works in noisy environments, or that higher noise requires more measurements, but do not provide any quantitative analysis to support these claims.

A good example of noise-resilience evaluation is~\cite{vicarte2022augury} which studies four noise sources on the memory-dependent prefetcher (DMP).
Likewise, \cite{oren2015spy} dedicates a full subsection to the sources, effects, and mitigations of noise on their attack. 

\paragraph{A6 Evaluated on a single platform}
20.7\% of papers evaluate their attacks on a single platform, usually a single CPU model.
This lack of diversity raises concerns about the generalizability of the results to other hardware configurations, and deters reproduction efforts.

However, it is noteworthy that some papers check if their attack could be applied to platforms other than the one tested, even without a full evaluation.
Others claim they did reproduce their attack on other platforms, but without giving any details about the results.
Lastly, some papers evaluate their attacks on multiple CPU models from the same vendor and the same microarchitecture generation, resulting in a limited diversity and relevance of the evaluation.

\cite{zhang2023bunnyhop, wang2022hertzbleed} demonstrate their attack by evaluating on multiple platforms from different vendors, showcasing the generalizability of their results.
Peculiar cases were treated conservatively, as described in \Cref{appendix:codebook}.

\paragraph{A7 Evaluated on a single configuration}
31.7\% of papers do not evaluate all of their attack primitives on multiple configurations.
Some papers only test their attacks under a single set of system settings, such as specific CPU frequencies, memory configurations, or network conditions.
This limits the relevance of the benchmark, as it does not show how the attack performs under different settings, nor how robust it is against noise, for instance.

As an example of good practice, \cite{yu2023synchronization} has 2 primitives, both tested under different workloads.
Likewise, \cite{pessl2016drama} evaluates the DRAM row-buffer-based leakage on various environments with both a covert channel and a side channel.

\subsection{B. Improper comparisons}

\paragraph{B1 Unfair comparisons to other work}
67.3\% of papers provide unfair comparisons. 
Most flawed papers compare their results to raw numbers of prior work obtained in different settings, hardware, or threat models.
In some cases, they cite other works but limit their comparison to either quantitative or qualitative data.

For 2 papers (3.8\%), we are unable to conclude on the fairness of their comparisons, due to their lack of clarity or completeness.
First, comparisons are made through unclear statements (\eg ``our results are similar to those in this paper'').
Second, they only compare parts of the properties (usually the ones at which the current work excels) but not all.

We would like to highlight that 2 papers~\cite{dutta2023spy, ahn2021networkonchip} explicitly acknowledge unfair comparisons (\eg due to differing setups).
We therefore do not label them as flawed.

A better practice would be to reproduce prior works under the same conditions to have a fair ground for comparison, as done in \cite{zhang2023mwait, tatar2022tlbdr}.

\paragraph{B2 No comparison to other work}
35.8\% of papers do not provide any comparison to prior work.
Often, this is due to papers putting their focus on demonstrating the mere \textit{feasibility} of an attack through toy examples, neglecting a full comparison.
For 1 paper, the evaluation's focus is on a novel mitigation to a side channel, leaving the evaluation of the attack itself with no comparison with prior work. 

Moreover, 33.3\% of papers only partially succeed in giving a comparison (let alone a \emph{fair} comparison).
This is mostly due to comparing only part of their evaluation, even when there is plenty of related work to compare to (\eg website fingerprinting).

For example, \cite{guo2022leaky} directly compare their approach against re-implementations of \PrimeProbe, \textsc{Prime+Scope}, and \textsc{Flush+\-Re\-load} in all evaluations.
Similarly, \cite{zhang2023mwait} implement \FlushReload and \PrimeProbe as baselines for comparison.

\paragraph{B3 Evaluation designed to highlight performances}
This flaw affects 16.9\% of papers.
Usually, this results from other flaws, such as A1 (no error rate); A6 (evaluation on a single platform); A7 (evaluated on a single configuration);  B1 (unfair comparison);  B2 (no comparison); B4 (benchmark only on highly controlled/simplified environments); C1 (missing target/target version) or C3 (no code/materials/artifact).

Still, other factors can lead to this flaw, like a toy example attack (an attack that is presented as a demonstration rather than a proper benchmark) mainly for showcasing performance.
In some cases, papers claim their primitive enables various side-channel attacks but do not actually fully demonstrate any attack.
We encountered one case where the attack was not carried out to completion: the paper recovered only a single byte and extrapolated the remaining ones.
Likewise, we found 1 paper that reports manual error correction, which is impractical and raises concerns about bias or correctness.
Some papers generalize the attack's applicability without experiments, or did covert-channel validation only on trivial message patterns (\eg sending only 1), or with the claim they can recover most of the key without providing any evidence on consistency, robustness, or reliability. 

For 1 paper (1.2\%), it is unclear whether the evaluation is designed to highlight performances or not, mostly because they claim to recover most of the key, with indication of where the missing bits are, but did not give any clue whether those results are consistent or even reliable.

Also, some works could initially be considered flawed, because of initial naive benchmarks or incomplete attack, but later provide strong benchmarks in a variety of contexts~\cite{tatar2022tlbdr} or can justify the incomplete attack~\cite{wang2023dvfs}.
We did not consider these papers as flawed as they were transparent about the limitations of the evaluation. 

\paragraph{B4 Benchmarks only on highly controlled/simplified systems}
16.9\% of papers only benchmark their attacks in highly controlled or simplified environments.
Indeed, many experiments depend on fine-tuned parameters, specific platform settings, or non-default system configurations.
Plus, some attacks assume privileges or access unlikely in practice (\eg MSR, RAPL, disabling prefetchers).
Others sometimes need patched or specially instrumented victim implementations, which the paper does not clearly acknowledge.
We also encountered the case where the setup disables OS services, fixes CPU frequency, isolates cores, or otherwise removes noise, and sometimes justifies ignoring noise altogether.

On the attacks themselves, they might be limited or partially demonstrated (\eg key-recovery results cover only a few bits; covert-channel experiments use only simple, repeating messages; side-channel evaluations remain toy examples) to only show feasibility and not full exploitation.
Due to all those aspects, it is unclear how the attacks would perform in realistic environments.

Attacks that explicitly require a specific threat model, such as an adversarial OS to target a TEE, were not considered flawed.
Other examples of realistic environments can be found in \cite{chen2024prefetchx} which attack real-word targets such as the network traffic patterns of users.
Another example is \cite{aldaya2022hyperdegrade} which shows many different benchmarks on real data (with a Raccoon attack on TLS-DH key exchanges) on different Intel architectures.

\subsection{C. Missing information}

\paragraph{C1 Missing target/target version}
27.6\% of papers do not clearly identify targets and versions.
Most provide only partial information on the target, or omit the driver/OS/kernel versions when they are relevant to the attack.
Some refer to targets imprecisely (\eg MbedTLS ``last version'' or OpenSSL's ``AES T-table'') without giving exact version numbers. 
This makes reproduction difficult, as different versions may exhibit different vulnerabilities or behaviors.

Targets and their version should be clearly specified for all attacks in the paper, such as in \cite{purnal2022double, yan2019attack, oren2015spy}.

\paragraph{C3 No code/materials/artifact}
49.4\% of papers do not provide any code or material to reproduce their results.
This issue manifests itself in three ways.
First, there is no link to the code in the paper, nor any code on the web.
Second, there is no link in the paper, but a repository exists.
Third, there is a link in the paper pointing to a now inaccessible resource, and no other accessible repository exists.

For 24.1\% of papers, only part of the artifact is available.
Others might be missing, containing only a subset of what the paper describes (\eg demos or tools instead of full attacks) or “TBA.” for many years so far. 
The code might also be poorly documented, making it unclear how to reproduce the experiments.
As a consequence, because key artifacts are absent or incomplete, the experiments cannot be reproduced in full.

We consider that this flaw did not apply to papers that did not put a link to their repositories in their paper, but did went through AE \cite{liu2023sidechannel, chen2024gofetch}, or are waiting for stake-holders' approval \cite{tan2021invisible}. 

One of the clearest examples of good reproducibility practices can be found in \cite{kogler2023collide}, where the authors provide a comprehensive repository containing all the necessary code and documentation (including duration and warning) to reproduce their experiments.

\paragraph{C4 No number of samples (side channel)}
42.8\% of papers do not fully provide the number of samples used in their side-channel attacks.
Most of the time, there is no clear number of samples provided for all the attacks, or they only give execution times but not the sample count needed for success.
For some papers with toy-examples, because those are minimal demonstrations, they appear to work with a single trace if no samples are given, but this might not be the case.
We also encountered a paper where even when sample counts are mentioned, they relate only to observing an effect, not necessarily to completing a full attack.
With no information about the number of samples, it becomes difficult to assess the practicality and efficiency of a side-channel attack.

Papers such as \cite{xu2024exploitation, tan2021invisible, dai2022dont, rauscher2024idleleak} provide clear information about the number of samples used in their side-channel attacks, enhancing the reproducibility and credibility of their results.

\paragraph{C7 Missing error-rate computation / error correction}
32.4\% of papers fully or partially miss major details about error-rate computation or error correction.
Indeed, they do not explicitly explain how they compute error rate or the error correction technique they may use.
1 flawed paper justifies it by indicating that it wanted to show the conservative result (with the lower bounds).
None of the other flawed papers give any justification for this omission.
For those which partially miss information on error-rate computation or correction, it is mostly because they mention error rates but never specify how they are calculated.
In some cases, we see unusual correction techniques like “blocks shorter than 4 samples are errors” or an 8-bit framing rule given with no justification or methodology.
In others, we get no standard, systematic error-correction mechanism, or framing described in a reproducible way.
Other papers state what might cause errors and how to avoid them, but provide no quantitative measurement.

Good examples of error-rate computation or correction can be found in works like \cite{zhao2022binoculars, ren2021see, rauscher2024idleleak} which dedicate at least a paragraph or a footnote to their error computation or correction mechanism.

\section{Discussion} \label{sec:discussion}

Beyond identifying benchmarking flaws, we also highlight several tendencies observed in our corpus.
We examine various aspects of benchmarking practices, including:
\begin{paraenum}
  \item the evolution in the prevalence of flaws
  \item code and artifacts availability
  \item trends in the number of benchmarks, platforms, and configurations evaluated per paper
\end{paraenum}.
Our goal is to understand how these practices have evolved over time, and how they differ between the security and architecture communities.

However, these results must be interpreted with caution, as our corpus is imbalanced, with more security papers and a greater number of recent publications.
Due to the small number of architecture papers before 2020, our analysis primarily focuses on the period from 2020 to 2024, while keeping in mind the potential biases this may introduce.
Most of the figures associated with our claims appear in \cref{appendix:tendencies_figures}.

\subsection{Benchmarking practices evolution over time}

\begin{figure}[t]
    \centering
    \includegraphics[width=\columnwidth]{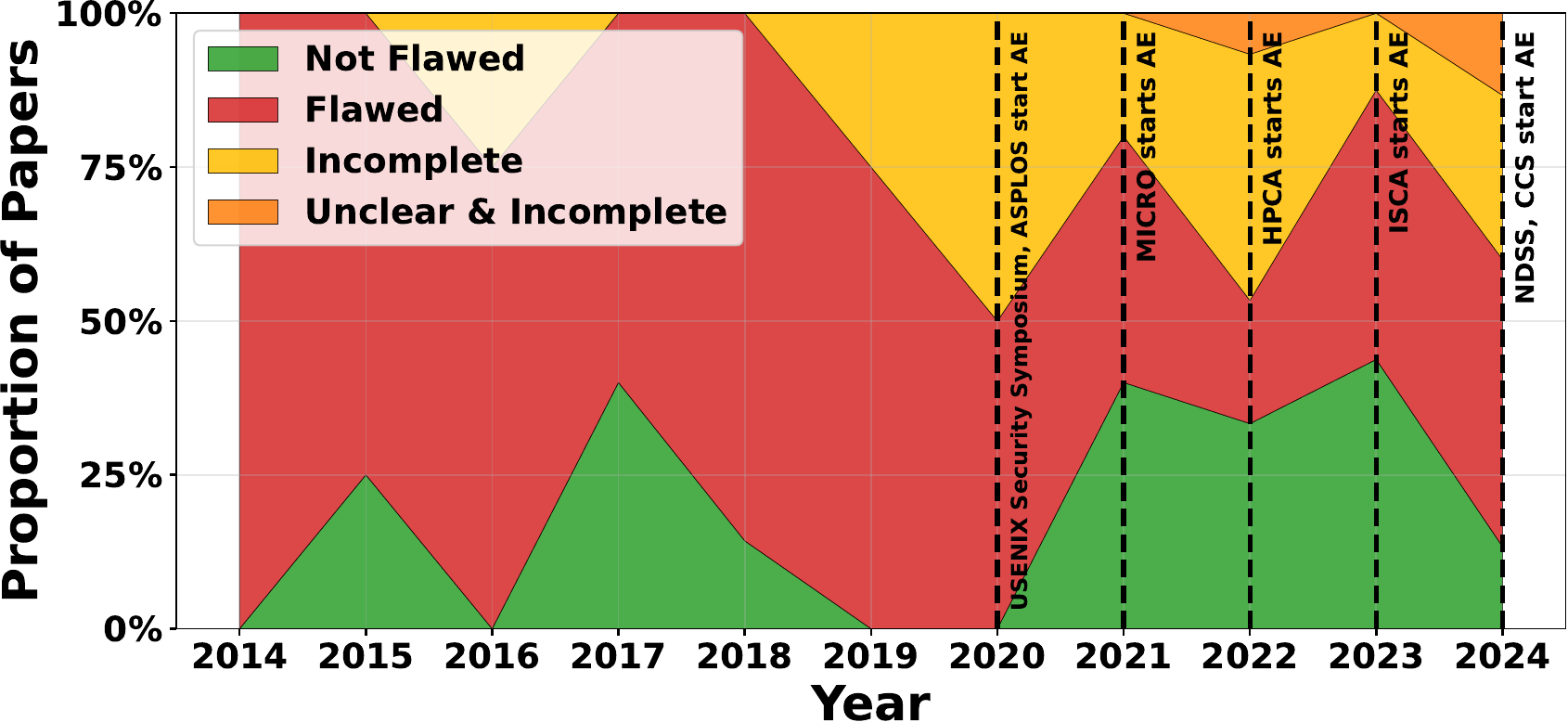}
    \caption{Availability of code, materials, and documentation over time for all conferences (with AE start).}
    \label{fig:porportion_plot_C3_all}
\end{figure}

\paragraph{Artifact evaluation} 
The adoption of AE by more conferences appears to have contributed to the improvement in availability of code and documentation.
These improvements are clearly visible in \cref{fig:porportion_plot_C3_all}, as the proportion of flawed papers decreased.
However, this problem is far from fully resolved, and many papers also remain partially flawed, with only parts of the evaluation code made available.
AE practices and expectations also vary from one venue to another, which would be interesting to study in future work.

\paragraph{General trends}
We observe a slight increase in the number of platforms evaluated per paper (\cref{fig:boxplot_number_of_benchmarked_platforms_per_year}).
More generally, we note that including a wide array of platforms (sometimes 10+) in the evaluation has become a well-anchored practice in recent publications.
The total number of benchmarks used for evaluating primitives increased after 2020 (\cref{fig:boxplot_number_of_evaluations_per_year}), primarily driven by a rise in benchmarks based on side-channel attacks.
This may reflect reviewing practices becoming stricter for this type of paper.
Conversely, interest in covert channels has declined slightly at the same time. 
However, we see no such trends in the number of benchmarked configurations (\cref{fig:boxplot_number_of_benchmarked_configuration_per_year}), with values varying considerably. 

\subsection{Communities' practices}

\paragraph{General trends} 
The number of flaws per paper reveals clear differences across communities.
Papers published in security venues contain fewer flaws overall (\cref{fig:boxplot_flawed_per_year_security}).
They seem to show an improvement since 2020, coinciding with the introduction of AE at the USENIX Security Symposium.
In contrast, architecture papers appear to exhibit more flaws (\cref{fig:boxplot_flawed_per_year_microarchitecture}), possibly reflecting differing expectations from reviewers and the community on evaluation.
For example, we noted that several papers provided limited details on the attack they implemented, or relied on toy examples, primarily to demonstrate feasibility.
We do not see the same trend of improvement for architecture conferences, though this may be due to the lower number of papers included overall.

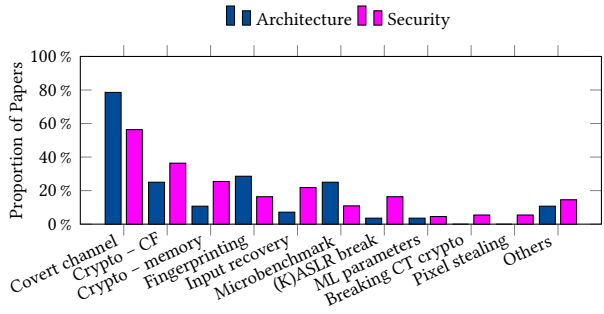
\begin{figure}[t]
\centering
\begin{tikzpicture}

\begin{axis}[
    ybar,
    bar width=6pt,
    width=\linewidth,
    height=3.8cm,
    ymin=0,
    ymax=100,
    ylabel={Proportion of Papers},
    y label style={yshift=-5pt, font=\footnotesize},
    yticklabel=\pgfmathprintnumber{\tick}\,\%,
    y tick label style={font=\footnotesize},
    symbolic x coords={
        Covert channel,
        Crypto -- CF,
        Crypto -- memory,
        Fingerprinting,
        Input recovery,
        Microbenchmark,
        (K)ASLR break,
        ML parameters,
        Breaking CT crypto,
        Pixel stealing,
        Others
    },
    xtick=data,
    x tick label style={rotate=25,anchor=east,font=\footnotesize},
    legend style={at={(0.5,1.1)}, anchor=south, legend columns=2, draw=none, font=\footnotesize},
    cycle list={{
        {fill={rgb,255:red,0;green,76;blue,153}}, 
        {fill={rgb,255:red,255;green,0;blue,255}}  
    }},
]

\addplot coordinates {
    (Covert channel,78.57)
    (Crypto -- CF,25.00)
    (Crypto -- memory,10.71)
    (Fingerprinting,28.57)
    (Input recovery,7.14)
    (Microbenchmark,25.00)
    ((K)ASLR break,3.57)
    (ML parameters,3.57)
    (Breaking CT crypto,0)
    (Pixel stealing,0)
    (Others,10.71)
};

\addplot coordinates {
    (Covert channel,56.36)
    (Crypto -- CF,36.36)
    (Crypto -- memory,25.45)
    (Fingerprinting,16.36)
    (Input recovery,21.82)
    (Microbenchmark,10.91)
    ((K)ASLR break,16.36)
    (ML parameters,4.55)
    (Breaking CT crypto,5.45)
    (Pixel stealing,5.45)
    (Others,14.55)
};

\legend{Architecture, Security}

\end{axis}
\end{tikzpicture}
\caption{Distribution of papers with covert channels and side-channel attack types (Architecture vs. Security conferences).}
\label{fig:side-channel_attack_types_per_conference_group}
\end{figure}

\paragraph{Evaluation}
Trends across communities are particularly visible in \textit{how} evaluations are conducted.
As shown in \cref{fig:side-channel_attack_types_per_conference_group}, security papers tend to evaluate their side-channel primitives mostly using a broad diversity of side-channel attacks.
Architecture papers instead primarily use covert channels for evaluation: 78.6\% evaluate at least one, compared to 56.4\% for security papers.
Security papers however tend to evaluate more platforms compared to architecture papers, as a way to demonstrate the robustness and applicability of their attacks (\cref{fig:boxplot_number_of_benchmarked_platforms_per_year}).
Both communities rely on similar metrics for covert-channel assessment (\cref{fig:covert_percentage_per_conf_group}): speed and error rate remain the dominant choices.
However, architecture papers more readily introduce specialized or original metrics (\eg cache miss rate, stability).
The increase in the number of benchmarks used for evaluation is mostly due to security papers (\cref{fig:boxplot_number_of_evaluations_per_year_security}) and architecture papers to a lesser extent (\cref{fig:boxplot_number_of_evaluations_per_year_microarchitecture}), although this trend must be nuanced given our small corpus.

\paragraph{Code availability}
Papers in security venues tend to share code substantially more often, although sometimes incompletely (\cref{fig:porportion_plot_C3_security}).
By contrast, even after the introduction of AE, more than half of architecture papers do not publish their code (\cref{fig:porportion_plot_C3_architecture}).

\section{Recommendations} \label{sec:recommandations}

Based on the previous sections, we provide recommendations on what to test when benchmarking a primitive.
A good benchmark depends not only on authors but also on reviewers, program committees, and the community itself.
We give recommendations for each group below.

\subsection{To authors}

\paragraph{Share the Code, Strengthen the Science}
Sharing artifacts such as code is crucial for advancing science and establishing the credibility of research.
It can clarify or supplement information in papers, facilitating a general understanding of experiments and results.

When authors provide access to the underlying code and materials used in their research, they enable other researchers to reproduce their results, validate their findings, and build upon their work.
This transparency fosters collaboration and accelerates the pace of scientific discovery, dissemination of knowledge, and best practices.
Furthermore, sharing code and materials enables the identification and correction of errors, resulting in more robust and reliable research outcomes.
However, if sharing the code would be unethical (\eg if mitigations are unavailable or impossible), the author should explicitly state this, as in \cite{tan2021invisible}.

\paragraph{Resolution is the Real Baseline}
This recommendation stems from flaws \emph{A2} and \emph{A3}.
So far, there is no consensus on how a new covert-channel or side-channel attack should be benchmarked.
Many studies focus on demonstrating attacks against specific cryptographic algorithms or applications (\eg AES T-table, modular exponentiation, ECDSA, (K)ASLR).
However, these attacks are merely proxy metrics for the real capabilities of their primitive and can be misleading.
They do not indicate how the primitive would perform in other scenarios or applications.

Generally, the greater the spatial resolution and the lower the temporal resolution are, the more impactful the primitive will be.
These metrics provide a clear understanding of the primitive's capabilities and limitations, allowing for meaningful comparisons with existing work and in accordance with the recommendations from \cite{SIGPLANEmpiricalEvaluationChecklist}.
Authors should also discuss the applicability of their primitive based on its resolution, highlighting effective and ineffective scenarios.
In the case where spatial or temporal resolution is not applicable, authors should clearly explain why and provide alternative metrics that better capture the primitive's performance.
In any case, replacing proxy metrics with resolution metrics improves the benchmark's completeness.

\paragraph{Fair Comparisons Build Trust}
This recommendation stems from flaws \emph{B1}, \emph{B2}, \emph{B3}, and \emph{B4}.
When benchmarking a new primitive, it is crucial to provide fair and honest comparisons to prior work whenever possible.
This involves re-implementing existing methods under the same or similar conditions as the new primitive to ensure a level playing field for comparison.
The introduction of a standardized common benchmark, or benchmark protocol, would help provide fair evaluations.
This could take the form of a framework helping authors to implement primitives, similarly to Mastik~\cite{MastikFramework}. 
If re-implementation is not feasible, authors should clearly acknowledge the limitations of their comparisons and avoid cherry-picking favorable results.

Additionally, authors should strive to evaluate their primitive in realistic environments that reflect practical scenarios.
This includes considering various system configurations, noise levels, and access privileges that an attacker might realistically have.
Providing fair comparisons and realistic evaluations builds trust in the results and contributes to the advancement of the field by demonstrating the relevance and soundness of the work.

\paragraph{Errors Should Never Pass Silently}
This recommendation stems from flaws \emph{A1}, \emph{A4}, \emph{B3}, and \emph{C7}.
When benchmarking a new primitive, it is essential to thoroughly evaluate and report error rates and noise resilience.
Authors should provide clear metrics on the reliability of their results, including error rates, accuracy, and related metrics.

Additionally, authors should assess how their primitive performs in the presence of noise, and quantify its impact on the results.
This includes evaluating the primitive under various noise conditions and discussing any mitigation strategies employed.

Furthermore, authors should detail their error-rate computation methods and any error correction techniques employed.
Transparent reporting of errors and noise resilience enhances the credibility of the benchmarks and provides valuable insights into the robustness of the primitive.
This transparency improves the completeness, relevance, soundness, and reproducibility of the benchmarks.

\paragraph{Generalization is not a Given}
This recommendation stems from flaws \emph{A6}, \emph{A7}, and \emph{B4}.
When benchmarking a new primitive, it is crucial to evaluate its performance across multiple platforms and configurations.
This includes testing the primitive on different hardware models, OS, and system settings to assess its generalization.

Authors should also consider evaluating their primitive in various environmental conditions, such as different noise levels and access privileges.
Demonstrating the robustness of their primitive across diverse scenarios provides a more comprehensive understanding of its capabilities and limitations.

Additionally, authors should clearly discuss the applicability of their primitive, highlighting scenarios in which it performs well and those in which it may struggle.
This transparency helps readers assess the completeness, the practical relevance, and the soundness of the primitive.

\paragraph{Explicit is Better Than Implicit}
This recommendation stems from flaws \emph{C1}, \emph{C4}, \emph{C7}, \emph{A2}, and \emph{A3}.
Flaws \emph{C2}, \emph{C5}, \emph{C6}, and \emph{C8} are also addressed, even though they were less prevalent in our corpus.
When benchmarking a new primitive, it is essential to provide explicit and detailed information about all relevant aspects of the evaluation.
This includes clearly specifying the target system and its version, the number of samples used in side-channel attacks, and the methods for error-rate computation and correction.

When possible, authors should provide comprehensive details on spatial and temporal resolution metrics.
Authors should include any other pertinent information, such as classification details, covert-channel protocol specifics, and prerequisites for the attack.
Being explicit about these details enhances the completeness, relevance, soundness, reproducibility, and general credibility of benchmarks, enabling others to accurately assess and build upon the work.

\paragraph{Every Claim Must Be Evaluated Properly}
This recommendation stems from flaw \emph{A5}, particularly applied to properties of stealthiness and detectability.
Although none of the papers in our corpus exhibited this flaw, we believe it is an important consideration.
If authors claim that their primitive is stealthy or difficult to detect, they must provide empirical evidence to support their assertion.
This may involve conducting experiments to assess the primitive's detectability under various conditions and comparing the results to those of existing methods.
Rigorously evaluating all claims strengthens the completeness and relevance of benchmarks and provides a more comprehensive understanding of a primitive's capabilities.

\subsection{To reviewers and program committees}

Beyond authors, reviewers in program committees play a crucial role in ensuring the quality of benchmarks and their evolution over time.
They critically evaluate the benchmarks presented in submitted papers and provide constructive feedback to authors.
We provide the following recommendation:

\paragraph{Encourage Relevant Benchmarks}
This recommendation stems from flaws \emph{A2}, \emph{A3}, \emph{A6}, \emph{A7}, and \emph{B3}.
Without micro-managing authors' work, reviewers and program committees should encourage authors to focus on relevant benchmarks that provide meaningful insights into the capabilities of their primitives.
Although end-to-end attacks are always appreciated, architecture papers should not prioritize them over a complete and thorough evaluation.
Instead, reviewers should advocate for benchmarks that assess the fundamental properties of primitives themselves, including spatial and temporal resolution metrics, and should discourage the use of proxy metrics that may not accurately reflect the primitive's performance.
Reviewers should also encourage the use of benchmarks that evaluate primitives in realistic environments and across multiple platforms and configurations.
Promoting relevant benchmarks helps improve the completeness, relevance, soundness, reproducibility, and general quality of research in the field.

\subsection{To the community}

Last but not least, the community has a significant role in shaping benchmarking practices.
The more the community emphasizes the importance of proper benchmarking, the more likely authors, reviewers, and program committees will prioritize it.
We provide the following recommendations:

\paragraph{Establish Standardized AE}
This recommendation stems from flaw \emph{C3}.
The community should work towards establishing standardized AE processes across conferences and journals, with clear guidelines and criteria to evaluate the reproducibility of research artifacts, such as code, materials, and documentation.
As of 2025, all major conferences in our corpus have adopted AE, though many smaller conferences and journals have not, and even where AE exists, it is not applied to all papers.
Therefore, venues should consider making AE mandatory—or provide stronger incentives—to ensure that published work meets a baseline level of transparency and reproducibility.
Recently, ~\cite{DBLP:conf/acmrep/DEliaDGMPPPSSV25, vansteenhuyse2026not, DBLP:conf/acmrep/OlszewskiLCBLBT25, vahldiek2026reprodb} evaluated and compared AEs for different top-ranked security and system venues and proposed best practices for their standardization.

\paragraph{Promote Comprehensive Benchmarking Practices}
This recommendation stems from the overall findings of our study.
The community should promote comprehensive benchmarking practices that encompass all relevant aspects of evaluating new primitives.
This includes encouraging authors to assess spatial and temporal resolution, error rates, noise resilience, generalizability, and other relevant metrics.
It should also advocate for transparent reporting of all evaluation details, including target systems, sample counts, and error correction methods.
Fostering a culture of comprehensive benchmarking will help ensure that research in the field is robust, reliable, and meaningful.

\section{Related Work}\label{sec:related_work}

Recent surveys on microarchitectural side-channel attacks by Ge \textit{et al.}~\cite{DBLP:journals/iacr/GeYCH16} and Lou \textit{et al.}~\cite{DBLP:journals/csur/LouZJZ21} provide broad overviews of the field.
Ge \textit{et al.} focus on shared environments, particularly cloud settings with co-resident VMs, while Lou \textit{et al.} adopt a more environment-independent perspective.
Both categorize attack techniques, outline common mitigation strategies, and evaluate cryptographic libraries such as OpenSSL and GNU Crypto, with findings consistent with ours.
However, their emphasis lies on attack classes, targets, and countermeasures rather than on benchmarking practices.

Related meta-analyses exist in other fields.
Schloegel \textit{et al.}~\cite{DBLP:conf/sp/SchloegelBSBSCEBMH24} evaluated the reproducibility of fuzzing papers and their adherence to established guidelines. In our work, such pre-existing criteria do not exist yet.
Rauscher \textit{et al.}~\cite{RauscherFKG25} identified often overlooked evaluation properties in side-channel research (\eg hit-miss margins, topological scope, attack time, blind-spot, etc.), focusing on cache attacks similar to \FlushReload and \PrimeProbe.
In contrast, we adopt a broader meta-level perspective not tied to any specific attack types.
They consistently benchmark inter-keystroke timing, KASLR, and AES T-table attacks, while we advocate for consistent resolution metrics rather than reusing specific attack scenarios.

Van der Kouwe \textit{et al.}~\cite{KouweHABG19} and De Meulemeester \textit{et al.}~\cite{demeulemeesterHardwareCostEvaluation2025} are the closest to our work methodologically, although in different fields.
The former focused on benchmarking flaws in systems defense papers, deriving recommendations to improve benchmark quality and scientific rigor.
The latter focused on hardware-cost evaluation strategies, highlighted reporting deficiencies, and proposed a uniform evaluation methodology.
Our recommendations draw inspiration from the SIGPLAN empirical evaluation checklist~\cite{SIGPLANEmpiricalEvaluationChecklist}, adapted to microarchitectural side-channel attacks and their specific benchmarking flaws.

\section{Conclusion} \label{sec:conclusion}

In this work, we conducted a comprehensive survey of 83 papers and identified 19 benchmarking flaws that undermine the validity of microarchitectural side-channel and covert-channel evaluations.
No paper in our corpus is free of flaws, with an average of 5.72 issues per paper.
Based on these findings, we provide recommendations for authors, reviewers, program committees, and the broader community to improve benchmarking practices.

Our analysis highlights the need to rethink how evaluations are designed, conducted, and assessed.
In particular, we advocate wider adoption of resolution metrics, which provide an attack-agnostic way to characterize leakage.
When used rigorously, these metrics enable meaningful evaluation even for weaker or non–end-to-end attacks and can fully support a paper's empirical contribution.
Improving benchmarking is not solely the responsibility of authors: reviewers and program committees play a crucial role in enforcing sound methodology and clear reporting standards.
Reproducibility is equally important---sharing code, artifacts, and experimental material reduces the likelihood of flawed evaluations and increases the reliability of published results.
We hope that our work supports the community in moving toward more robust and transparent benchmarking practices for microarchitectural side-channel and covert-channel attacks and ultimately encourages progress toward a unified, widely accepted benchmark suite for this domain.

\section*{Acknowledgments}
This work benefited from the support of the AID, ANR-19-CE39-0007 MIAOUS, and ANR-21-CE39-0019/Deutsche Forschungsgemeinschaft (DFG) 491039149 FACADES.

\appendix

\bibliographystyle{ACM-Reference-Format}
\bibliography{bibliography}

\section{Open science}\label[appendix]{appendix:data}

All the material used for this study is available for consultation and reuse.
This includes the form used to fill in information for each article, and all the data we collected on the papers we selected for our study.
The code we used to analyze the data and produce the figures is also available on github \faicon{github}: \url{https://github.com/Naeele/Practice_Makes_-Im-Perfect---Artifacts} and Software Heritage:
\url{https://archive.softwareheritage.org/swh:1:dir:dcc278cbba073efa34c7fc75b06871c2a8720b9a;origin=https://github.com/Naeele/Practice_Makes_-Im-Perfect---Artifacts;visit=swh:1:snp:432e565b7676d5114147059b401f24cf2215f660;anchor=swh:1:rev:cf592adc274baeb2e935e60ac771cf64c41ad3d9}

\section{Ethical considerations} 
Our work raises no ethical concerns: it involves no human sub-
jects, user data, or real-world vulnerability testing, but only a meta-
analysis of existing research. The paper evaluation was conducted
over an extended period to avoid overburdening the authors.

\section{Code-book} \label[appendix]{appendix:codebook}

The following code-book was used to categorize articles presenting the same properties regarding their evaluation.
As contributions and authors may address similar concepts in different ways, it is useful to keep track of the reasoning behind the labeling of data.

\begin{itemize}
    \item \hypertarget{codebook:A1}{\textbf{\defref{A1}}} No error rate (covert channels or attacks with bit/byte decisions):
    \begin{itemize}
        \item \okKeyword: error rate reported.
        \item \unclearKeyword: error rate is implicit, but not clearly identified (\eg through the capacity).
        \item \incompleteKeyword: error rate reported only for some settings/attacks.
        \item \flawedKeyword: claims ``zero errors`` without measurement/proof, no claim regarding errors, or reports accuracy without a definition.
        \item \naKeyword: the paper does not claim an attack (\eg paper centered on improving a building block that does not affect error rate).
    \end{itemize}
    \item \hypertarget{codebook:A2}{\textbf{\defref{A2}}} No spatial resolution (what unit is distinguished: cache line, page, instruction/port, set/bank, core, etc.):
    \begin{itemize}
        \item \okKeyword: resolution unit is stated or deducible from a clear mapping (\eg reload threshold per cache line).
        \item \unclearKeyword: clues exist but mapping not explicitly stated.
        \item \incompleteKeyword: stated for one attack variant but not others evaluated.
        \item \flawedKeyword: resolution not mentioned.
        \item \naKeyword: the paper does not claim an attack where there are two units to distinguish spatially, or it is centered on improving a building block that does not impact error rate.
    \end{itemize}
    \item \hypertarget{codebook:A3}{\textbf{\defref{A3}}} No temporal resolution (sampling rate/event frequency distinguishable, doing a \FlushReload in measurable time):
    \begin{itemize}
        \item \okKeyword: authors report elements such as max sustainable sampling rate (or time per sample), or a clear statement of the time between measurement that can be achieved.
        \item \unclearKeyword: the temporal resolution is implicit, or we have some information that seems to contradict each other.
        \item \incompleteKeyword: authors report rate only for one config or reported as throughput without symbol timing.
        \item \flawedKeyword: rates inferred from unrelated timers (\ie relying only on reader experience) or not mentioned.
        \item \naKeyword: the paper does not claim an attack (\eg paper centered on improving a building block that does not affect error rate).
    \end{itemize}
    \item \hypertarget{codebook:A4}{\textbf{\defref{A4}}} No evaluation of noise resilience (co-runner noise, scheduler, interrupts, background load, network jitter, UI activity, etc.). This point makes no claim about the quality of the noise evaluation: only whether it is performed, and well reported.
    \begin{itemize}
        \item \okKeyword: measured performance under one or more realistic noise sources.
        \item \incompleteKeyword: only qualitative discussion, noise is mentioned and maybe tested a bit, but not in a meaningful manner.
        \item \flawedKeyword: claims ``robust to noise'' with no measurement, no mention of noise.
        \item \naKeyword: the paper does not claim an attack (\eg paper centered on improving a building block that do not impact error rate).
    \end{itemize}
    \item \hypertarget{codebook:A5}{\textbf{\defref{A5}}} No evaluation against claimed mitigations they bypass:
    \begin{itemize}
        \item \okKeyword: actually tests with the mitigation enabled (or faithful re-implementation) and measures effect.
        \item \incompleteKeyword: mitigation only discussed or tested partially.
        \item \flawedKeyword: draws bypass conclusion but does not evaluate against it.
        \item \naKeyword: no bypass claim. This is usually the case.
    \end{itemize}
    \item \hypertarget{codebook:A6}{\textbf{\defref{A6}}} Evaluated on a single platform:
    \begin{itemize}
        \item \okKeyword: at least 2 distinct platforms (\eg different µarch generations/vendors), or credible portability argument + partial replication.
        \item \flawedKeyword: only one platform (no platform also, but it's unlikely).
        \item \naKeyword: if the targeted mechanisms were only available on a restricted set of platforms at publication time.
    \end{itemize}
    \item \hypertarget{codebook:A7}{\textbf{\defref{A7}}} Evaluated on a single configuration (no parameter change/variants):
    \begin{itemize}
        \item \okKeyword: explores key parameters (thresholds, window sizes, eviction lengths, symbol times, stride sizes, core affinity, freq governors, etc.) for each primitive.
        \item \incompleteKeyword: Multiple primitives are presented, but only some are tested on multiple configurations.
        \item \flawedKeyword: single fixed configuration.
        \item \naKeyword: the paper does not claim an attack or provides more of theoretical claims (\eg paper centered on improving a building block that is not affected by configuration).
    \end{itemize}
    \item \hypertarget{codebook:B1}{\textbf{\defref{B1}}} Unfair comparisons to other work:
    \begin{itemize}    
        \item \okKeyword: normalizes for hardware/protocol, or re-implements baselines fairly.
        \item \unclearKeyword: comparison exists but insufficient detail to judge fairness (\eg no quantitative element, or unclear statements ("similar to etc.")).
        \item \flawedKeyword: raw number comparison across different protocols/hardware/threat models without normalization, cherry-picking weak baselines.
    \end{itemize}
    \item \hypertarget{codebook:B2}{\textbf{\defref{B2}}} Lack of comparison:
    \begin{itemize}
        \item \okKeyword: compares against prior closest work (metrics or at least qualitative rationale).
        \item \incompleteKeyword: not all evaluations are compared, or only partially (\eg only the covert channel and not any side channel).
        \item \flawedKeyword: no comparison.
        \item \naKeyword: genuinely first to demonstrate a new primitive/feature where no meaningful baseline exists.
    \end{itemize}
    \item \hypertarget{codebook:B3}{\textbf{\defref{B3}}} Evaluation designed to highlight performances:
    \begin{itemize}
        \item \okKeyword: uses realistic payloads or shows both easy and hard cases.
        \item \unclearKeyword: pattern choices not described.
        \item \flawedKeyword: uses trivial patterns (\eg long runs of 1s) or partial attacks but generalizes to full attacks without evidence.
    \end{itemize}
    \item \hypertarget{codebook:B4}{\textbf{\defref{B4}}} Benchmarks only on highly controlled/simplified systems:
    \begin{itemize}
        \item \okKeyword: includes realistic unprivileged scenarios or argues the necessity of controls with additional runs.
        \item \flawedKeyword: requires privileged instrumentation or constraints that invalidate stated threat model, only sandboxed/microbenchmarks with no real-system validation.
    \end{itemize}
    \item \hypertarget{codebook:C1}{\textbf{\defref{C1}}} Missing target/target version (OS, kernel, browser, driver, library version relevant to the attack):
    \begin{itemize}
        \item \okKeyword: clear reporting of versions relevant to the evaluation (library version for side channels targeting specific software, etc.).
        \item \incompleteKeyword: partial information (\eg OS but no version, attack on "AES T-table" without pointing to a specific implementation, etc.)
        \item \flawedKeyword: no version indication (microarchitecture excluded)
    \end{itemize}
    \item \hypertarget{codebook:C2}{\textbf{\defref{C2}}} Missing CPU/GPU model (model name \& µarch at minimum):
    \begin{itemize}
        \item \okKeyword: Clear report of the targeted platform
        \item \flawedKeyword: No model, or very broad statement when the behavior would vary from one target to the other (\eg "tested on a Skylake" without model)
    \end{itemize}
    \item \hypertarget{codebook:C3}{\textbf{\defref{C3}}} No code/materials/artifact (link or Artifact Evaluation (AE)). To avoid confusion, authors went through a shallow review of the artifact (reading documentation and browsing through the content). They did \textbf{not} try to reproduce the artifact, and AE badges are not factored in the annotation. While it would have been an interesting way to raise concerns and awareness toward the AE process (which may let some flaw pass), AE practices vary a lot in different venues.
    \begin{itemize}
        \item \okKeyword: material is available to reproduce all claims, and committed to the paper. An independent repository, not linked to the paper or an AE does \textbf{not} fall in this category.
        \item \unclearKeyword: material is available, but only with poor or no documentation.
        \item \incompleteKeyword: material is available, but only part of the contribution is made available.
        \item \flawedKeyword: material is not available, or available through an independent (\ie not committed to the paper) medium.
    \end{itemize}
    \item \hypertarget{codebook:C4}{\textbf{\defref{C4}}} No number of samples (side channel) (per class, per trace length, repetitions):
    \begin{itemize}
        \item \okKeyword: clear reporting of the amount of samples required to perform the attack.
        \item \unclearKeyword: numbers need to be extrapolated, or numbers correspond to unusual metrics that make it hard to compare.
        \item \incompleteKeyword: reports numbers only for some attacks
        \item \flawedKeyword: no number reported.
        \item \naKeyword: no side channel evaluated.
    \end{itemize}
    \item \hypertarget{codebook:C5}{\textbf{\defref{C5}}} Missing classification details (model type, features, windowing, hyper-parameters, train/val/test protocol):
    \begin{itemize}
        \item \okKeyword: gives enough details on the classification method and parameters to reproduce the classification.
        \item \incompleteKeyword: only give part of the parameters.
        \item \flawedKeyword: no information on the classifier.
        \item \naKeyword: no classification in the post-processing of traces.
    \end{itemize}
    \item \hypertarget{codebook:C6}{\textbf{\defref{C6}}} Missing covert-channel protocol (init/sync, unilateral/bilateral, framing, clocking, shared knowledge):
    \begin{itemize}
        \item \okKeyword: protocol is thoroughly described, including initialization, synchronization, transmission, unilateral/bilateral setting, etc.
        \item \unclearKeyword: information is confusing, or we need to extrapolate part of the protocol.
        \item \incompleteKeyword: part of the protocol is missing.
        \item \flawedKeyword: protocol is not described.
        \item \naKeyword: no covert channel evaluated.
    \end{itemize}
    \item \hypertarget{codebook:C7}{\textbf{\defref{C7}}} Missing error-rate computation / error correction (metric definition-bit/byte/word, alignment method, desync handling, edit distance vs Hamming, ECC scheme, and overhead):
    \begin{itemize}
        \item \okKeyword: error rate computation and potential ECC are described, including how they handle de-synchronization if relevant, how they correct errors if relevant. 
        \item \unclearKeyword: relevant information is not directly stated, but implicit in another metric.
        \item \incompleteKeyword: missing information.
        \item \flawedKeyword: no error rate reported.
        \item \naKeyword: no covert channel evaluated, no error rate reported, and no ECC used.
    \end{itemize}
    \item \hypertarget{codebook:C8}{\textbf{\defref{C8}}} Missing prerequisites / changing prerequisites across attacks (privileges, co-location, timers, hugepages, perf counters, freq scaling, page dedup, SMT, co-runner control):
    \begin{itemize}
        \item \okKeyword: clear reporting of prerequisites when they change, for each attack presented.
        \item \unclearKeyword: multiple prerequisites are introduced, but it is unclear which variant they are related to, some prerequisites are implicitly introduced.
        \item \flawedKeyword: prerequisites are not clearly stated, or contradictory.
        \item \naKeyword: single attack, or single prerequisite.
    \end{itemize}
\end{itemize}

\section{List of papers included in the study} \label[appendix]{appendix:papers}

The list can be found in \cref{tab:papers}.

\section{Usual targets} \label[appendix]{appendix:usual_targets}
For control-flow cryptographic attacks, we identified 41 attacks targeting various implementations, with some victims being targeted more frequently than others:

\begin{itemize}
    \item GnuPG v1.4.13 (7 times), v1.4.14, and 1 unknown version,
    \item GnuPG libgcrypt v1.6.3 (5 times), v1.5.2 (3 times), v1.5.0, and v1.7.5, 
    \item OpenSSL v1.1.0h (2 times), v1.0.2e, v1.0.1e, v1.1.1h, v3.2.
\end{itemize}

GnuPG and its associated library libgcrypt are the most commonly targeted victims in this category.

For fingerprinting attacks, we identified 23 attacks, with the following victims being the most frequently targeted:
\begin{itemize}
    \item Alexa top 100 websites (3 times),
    \item Chrome v63.0.3239.84 (2 times), v103.0.5060.114, and v121.0.
\end{itemize}
Alexa top 100 websites and various versions of the Chrome browser are the most commonly targeted victims in this category.
Other fingerprinting targets include Firefox, other known websites and famous applications from NVIDIA toolkit.
Out of the 23 attacks, 9 (39.1\%) target unknown or unspecified victims.

For memory access-based cryptographic attacks, we identified 17 attacks, with the following victims being the most frequently targeted:
OpenSSL v1.0.1e (5 times), v1.0.1f (2 times), v0.9.8, v0.9.8b, v1.1.0g, v1.0.2, and 2 unknown versions.
OpenSSL (and specifically OpenSSL v1.0.1e) is the most commonly targeted victim in this category.

It is also worth mentioning the diversity of software libraries, compiler versions, and SSL/TLS libraries used in the surveyed papers.
For instance, we observed the following:
\begin{itemize}
    \item MbedTLS versions: v1.3.10, v2.5, v2.6, v2.7, v2.13.0, v2.14, v2.15, v2.16, v2.28, v3.0.0, v3.0 (2 times), v3.1, v3.4.0, and 1 unknown version,
    \item GCC versions: v7.5, v8.4, v9.4, v10.3,
    \item WolfSSL versions: v4.2.0, v5.6.4.
\end{itemize}

\section{Tendencies' figures} \label[appendix]{appendix:tendencies_figures}
In this section, the reader can find the figures related to the tendencies described in \cref{sec:discussion}. These figures were not included in the main body of the paper for space reasons.

\begin{figure}[h]
    \centering
    \includegraphics[width=\columnwidth]{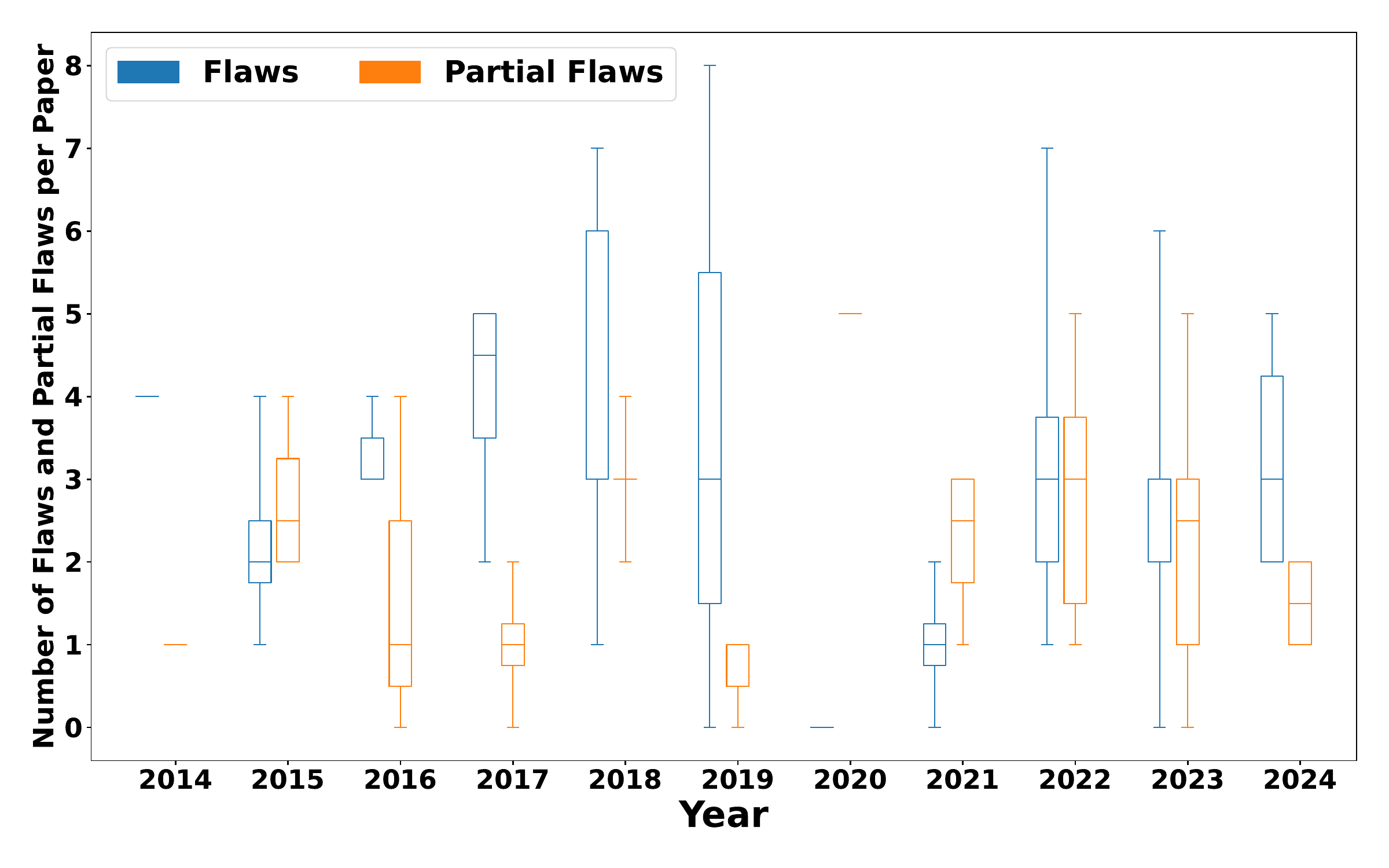}
    \caption{Boxplot of the number of flaws per paper per year, for security conferences.}
    \label{fig:boxplot_flawed_per_year_security}
\end{figure}

\begin{figure}[h]
    \centering
    \includegraphics[width=\columnwidth]{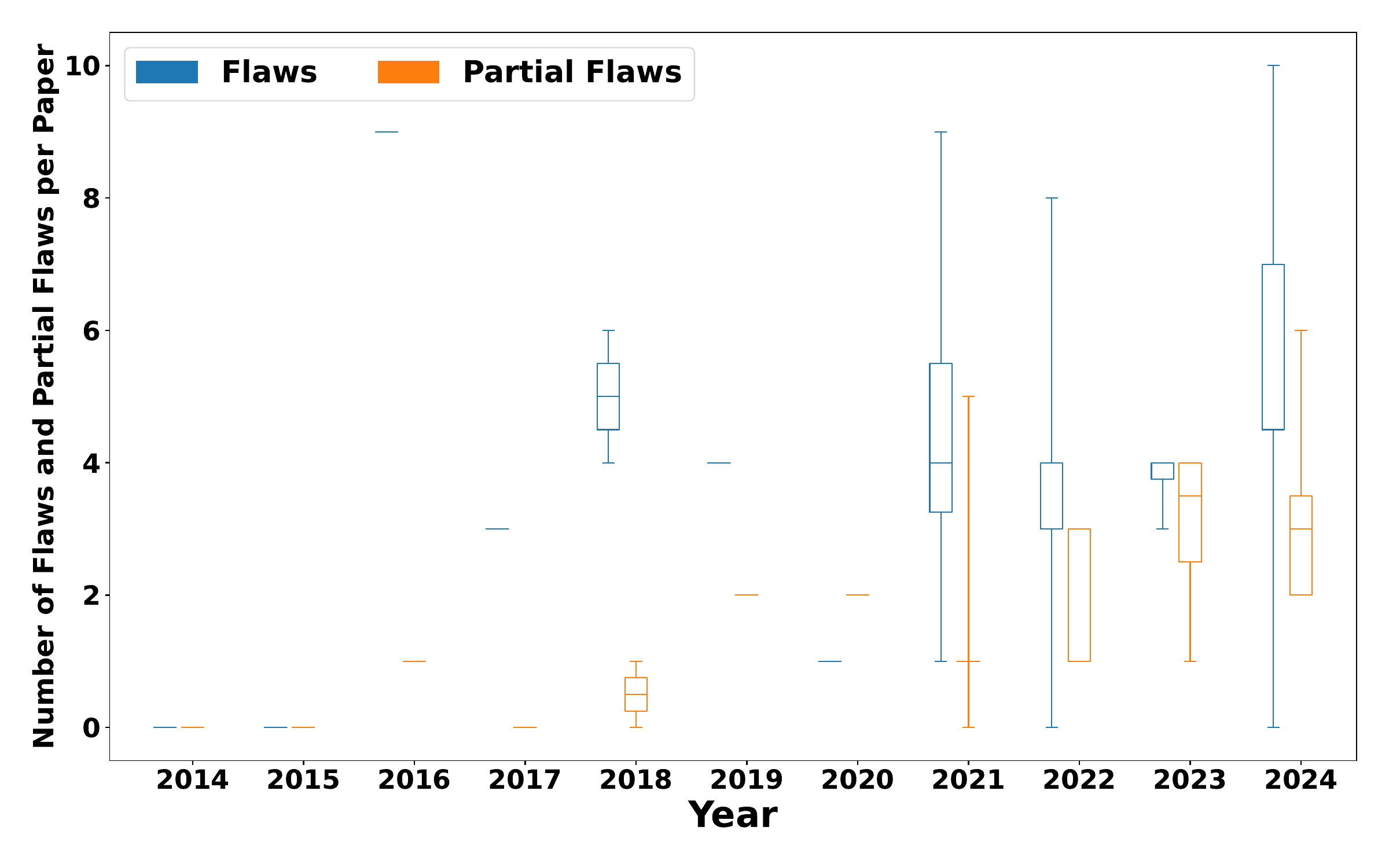}
    \caption{Boxplot of the number of flaws per paper per year, for architecture conferences.}
    \label{fig:boxplot_flawed_per_year_microarchitecture}
\end{figure}

\begin{figure}[h]
    \centering
    \includegraphics[width=\columnwidth]{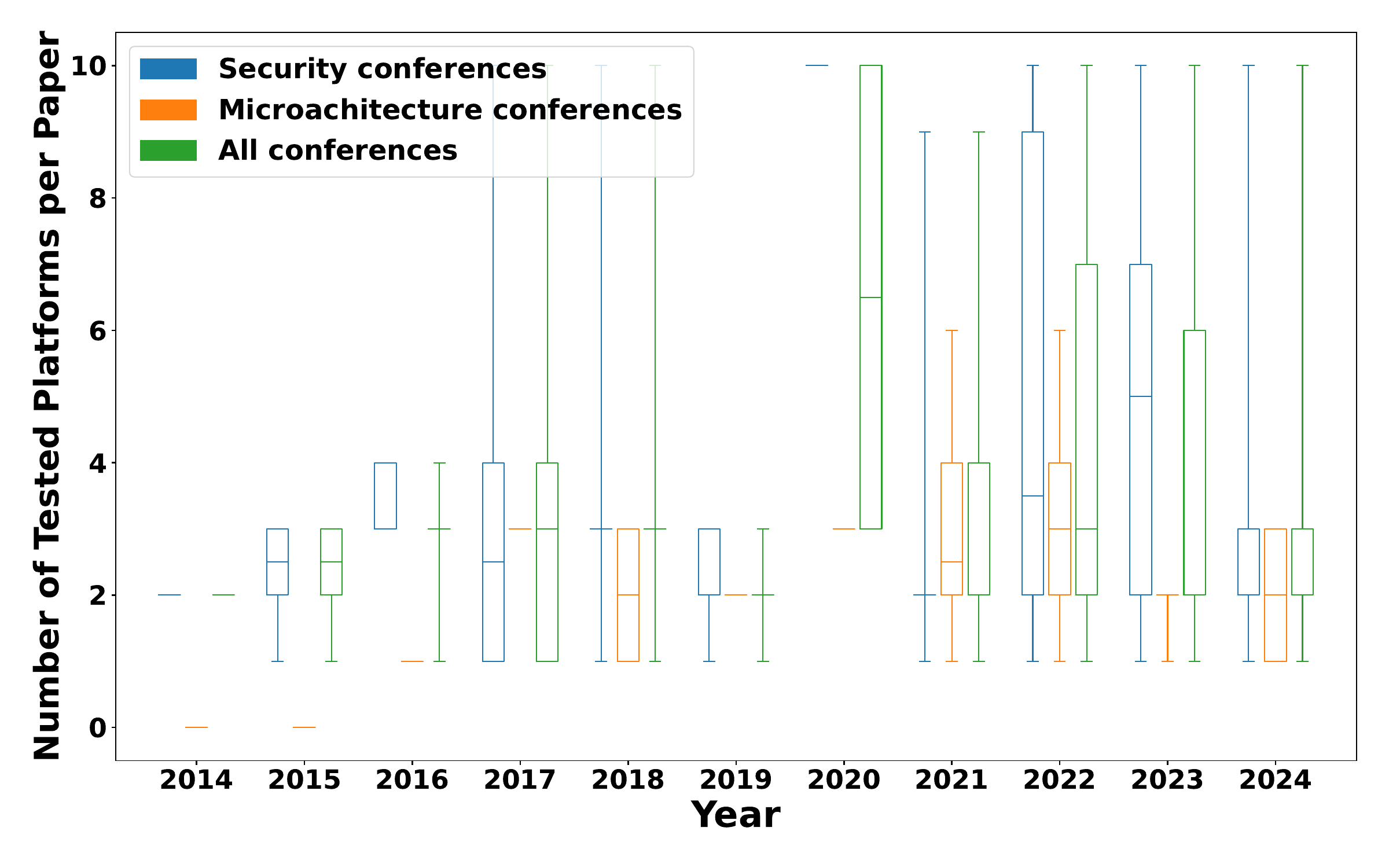}
    \caption{Boxplot of number of benchmarked platforms per paper per year, for all conferences.}
    \label{fig:boxplot_number_of_benchmarked_platforms_per_year}
\end{figure}

\begin{figure}[h]
    \centering
    \includegraphics[width=\columnwidth]{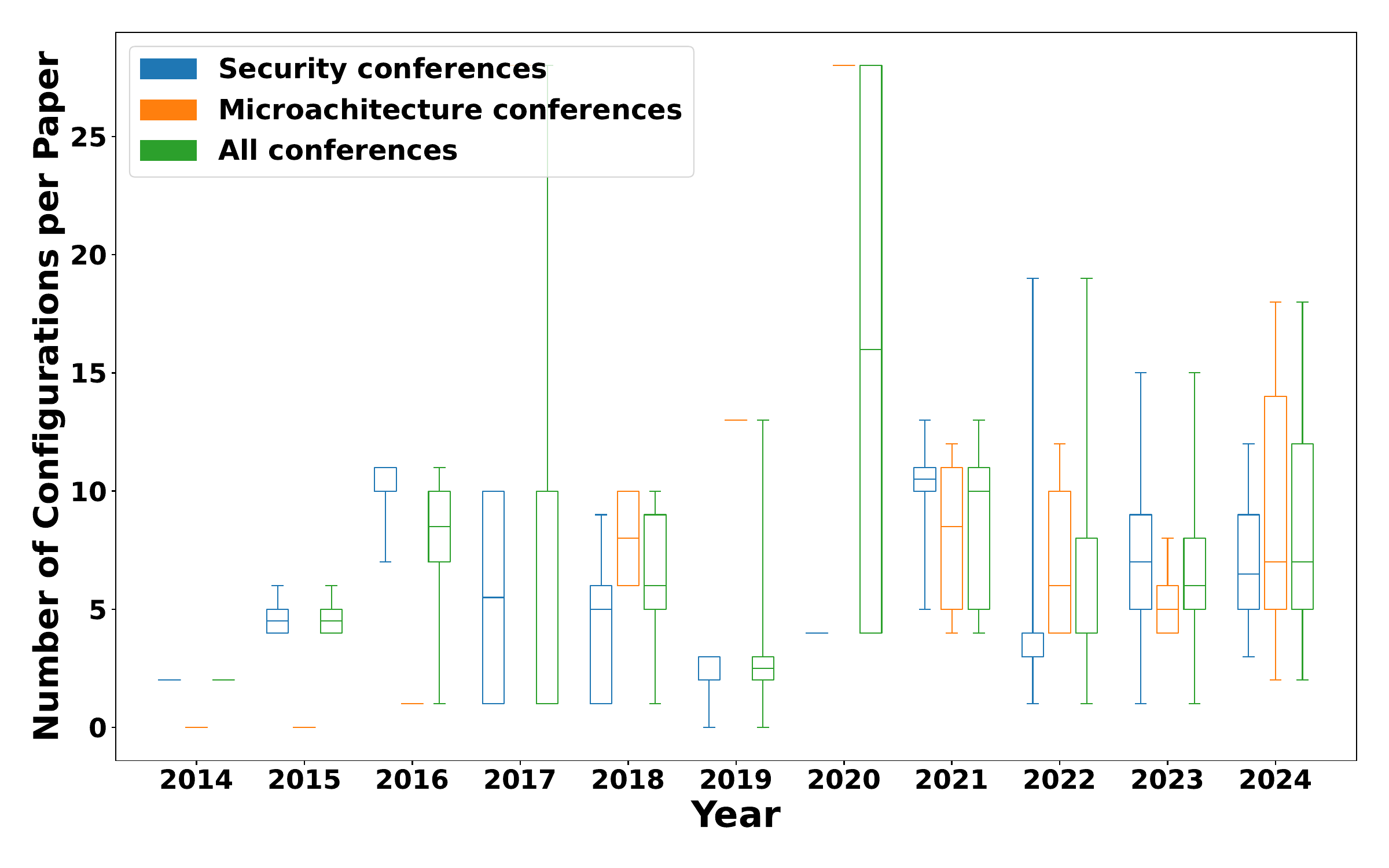}
    \caption{Boxplot of number of benchmarked configurations per paper per year, for all conferences.}
    \label{fig:boxplot_number_of_benchmarked_configuration_per_year}
\end{figure}

\newpage
\begin{figure}[h]
    \centering
    \includegraphics[width=\columnwidth]{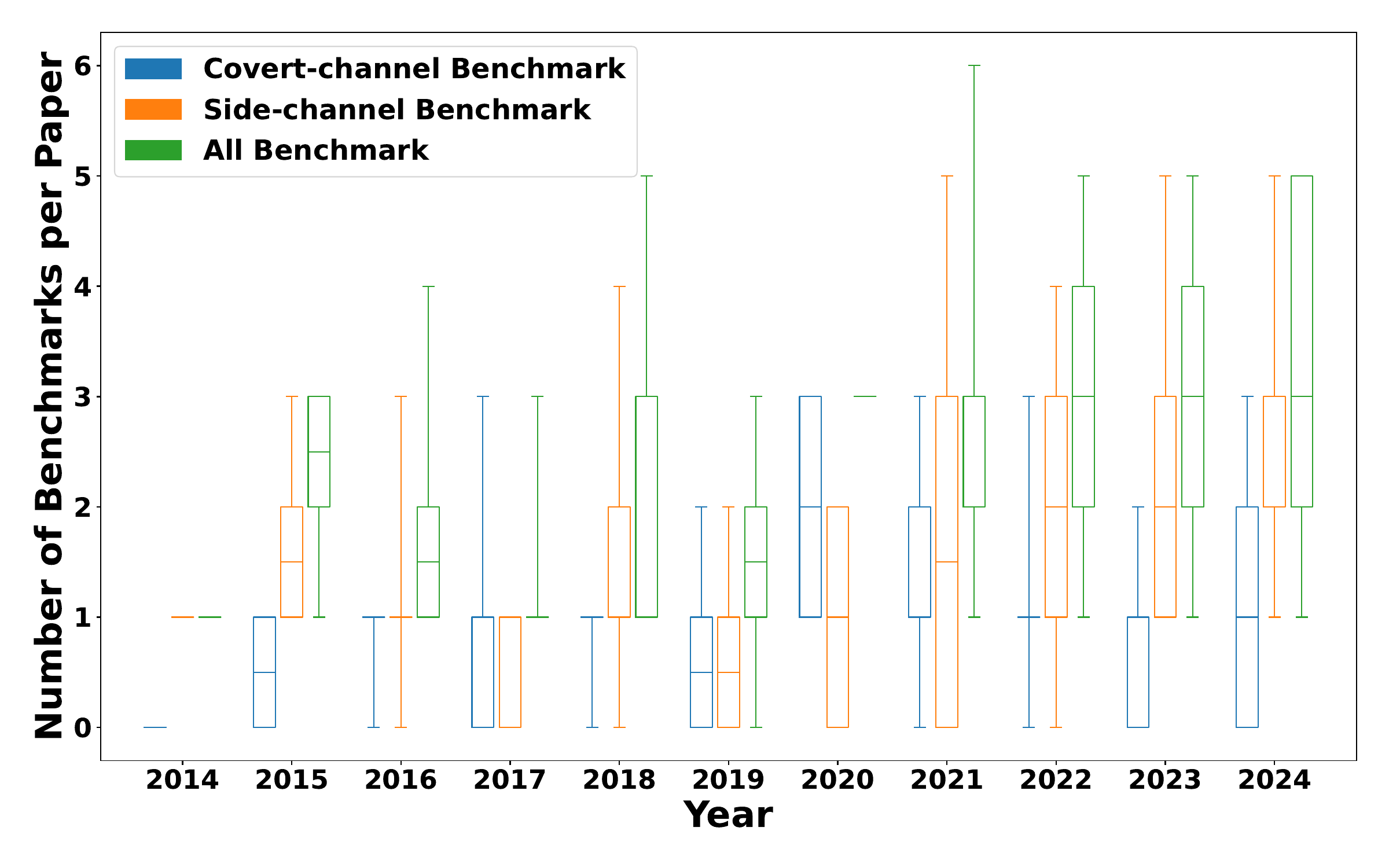}
    \caption{Boxplot of the number of evaluations per paper per year, for all conferences.}
    \label{fig:boxplot_number_of_evaluations_per_year}
\end{figure}

\begin{figure}[h]
    \centering
    \includegraphics[width=\columnwidth]{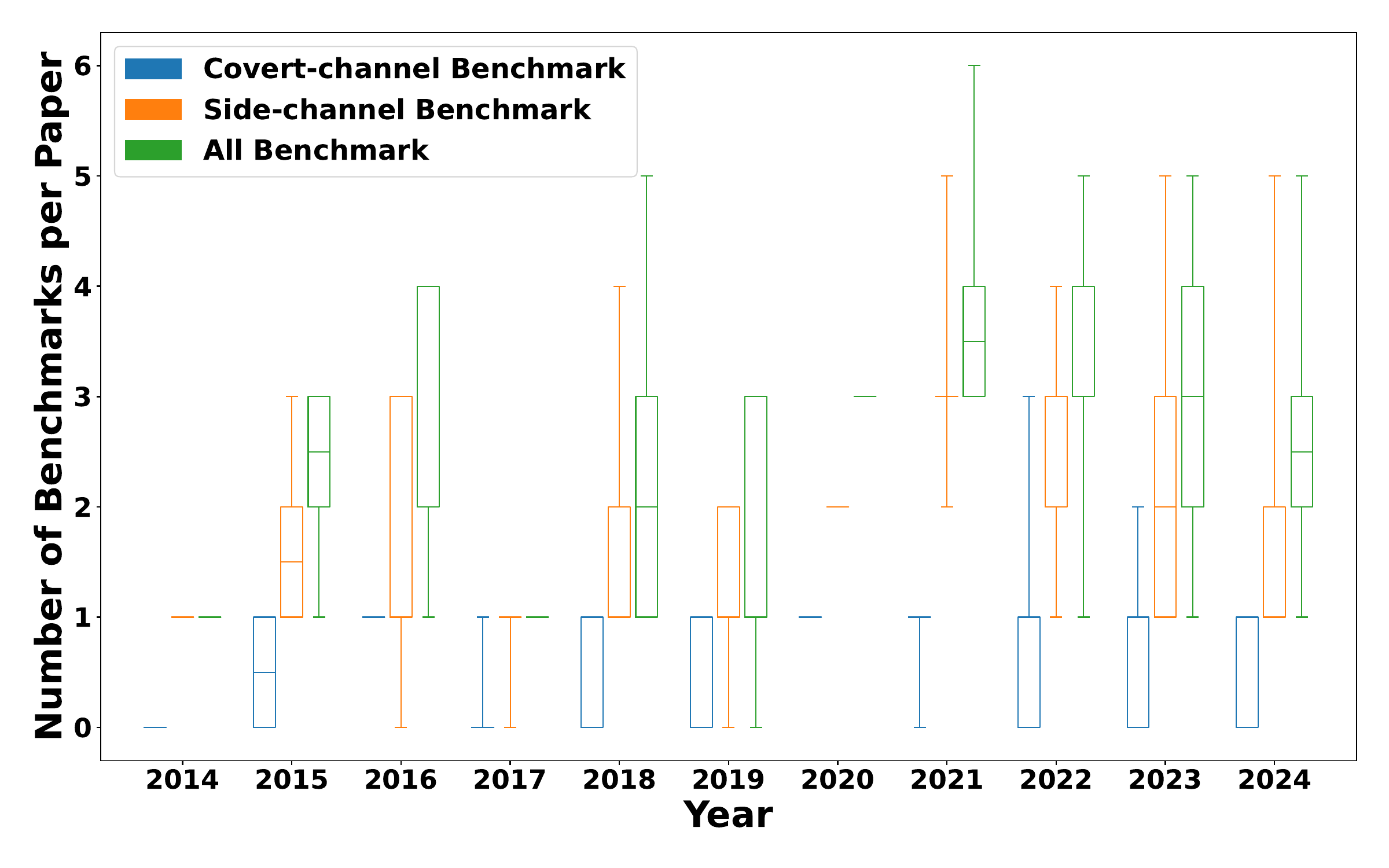}
    \caption{Boxplot of the number of evaluations per paper per year for security conferences.}
    \label{fig:boxplot_number_of_evaluations_per_year_security}
\end{figure}

\newpage
\begin{figure}[h]
    \centering
    \includegraphics[width=\columnwidth]{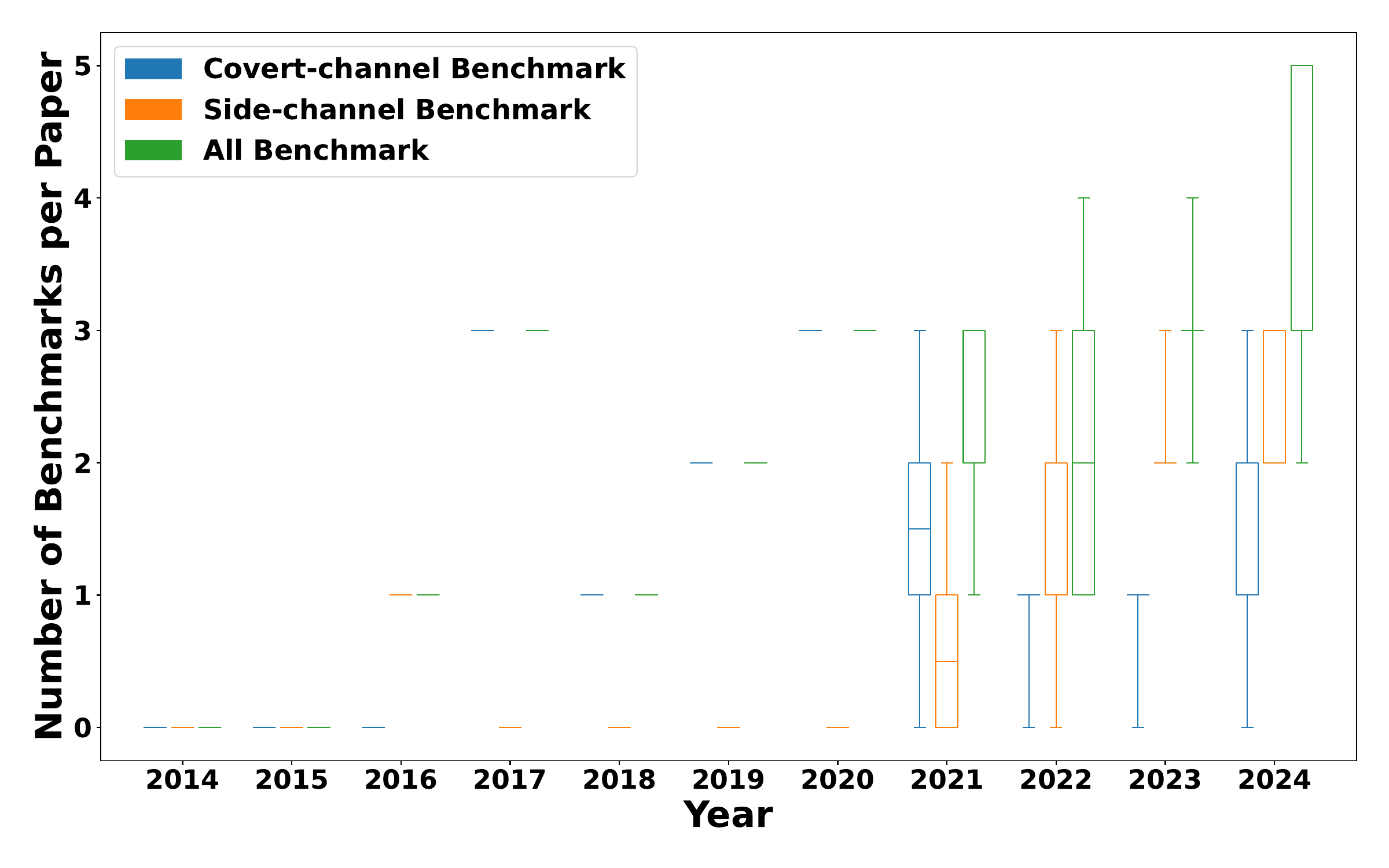}
    \caption{Boxplot of the number of evaluations per paper per year, for architecture conferences.}
    \label{fig:boxplot_number_of_evaluations_per_year_microarchitecture}
\end{figure}

\newpage
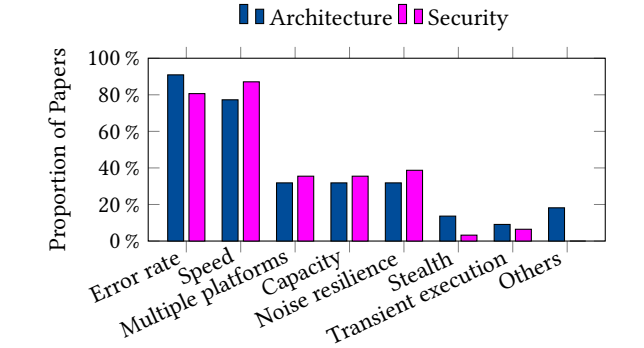
\begin{figure}[h]

\centering
\begin{tikzpicture}

\begin{axis}[
    ybar,
    bar width=6pt,                 
    width=0.9\linewidth,          
    height=4cm,                  
    ymin=0,
    ymax=100,
    ylabel={Proportion of Papers},
    yticklabel=\pgfmathprintnumber{\tick}\,\%,
    symbolic x coords={
        Error rate,
        Speed,
        Multiple platforms,
        Capacity,
        Noise resilience,
        Stealth,
        Transient execution,
        Others
    },
    xtick=data,
    x tick label style={rotate=25,anchor=east},
    legend style={at={(0.5,1.1)}, anchor=south, legend columns=2, draw=none},
    cycle list={{
        {fill={rgb,255:red,0;green,76;blue,153}}, 
        {fill={rgb,255:red,255;green,0;blue,255}}  
    }},
]

\addplot coordinates {
    (Error rate,90.91)
    (Speed,77.27)
    (Multiple platforms,31.82)
    (Capacity,31.82)
    (Noise resilience,31.82)
    (Stealth,13.64)
    (Transient execution,9.09)
    (Others,18.18)
};

\addplot coordinates {
    (Error rate,80.65)
    (Speed,87.10)
    (Multiple platforms,35.48)
    (Capacity,35.48)
    (Noise resilience,38.71)
    (Stealth,3.23)
    (Transient execution,6.45)
    (Others,0)
};

\legend{Architecture, Security}

\end{axis}
\end{tikzpicture}

\caption{Distribution of covert-channel reported properties among papers (Architecture vs. Security conferences).}
\label{fig:covert_percentage_per_conf_group}
\end{figure}

\begin{figure}[h]
    \centering
    \includegraphics[width=\columnwidth]{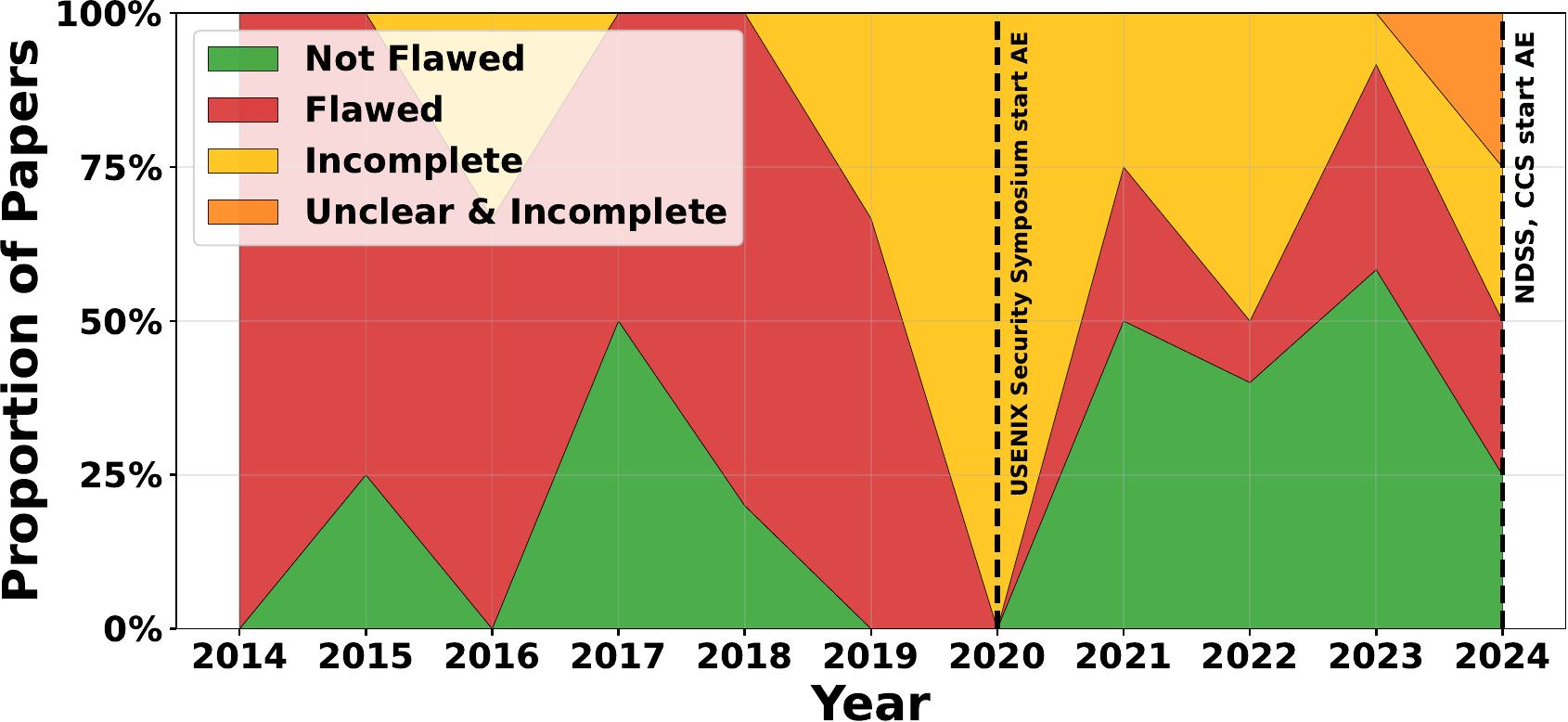}
    \caption{Availability of code, materials, and documentation for papers in security conferences over time (with AE start).}
    \label{fig:porportion_plot_C3_security}
\end{figure}

\begin{figure}[h]
    \centering
    \includegraphics[width=\columnwidth]{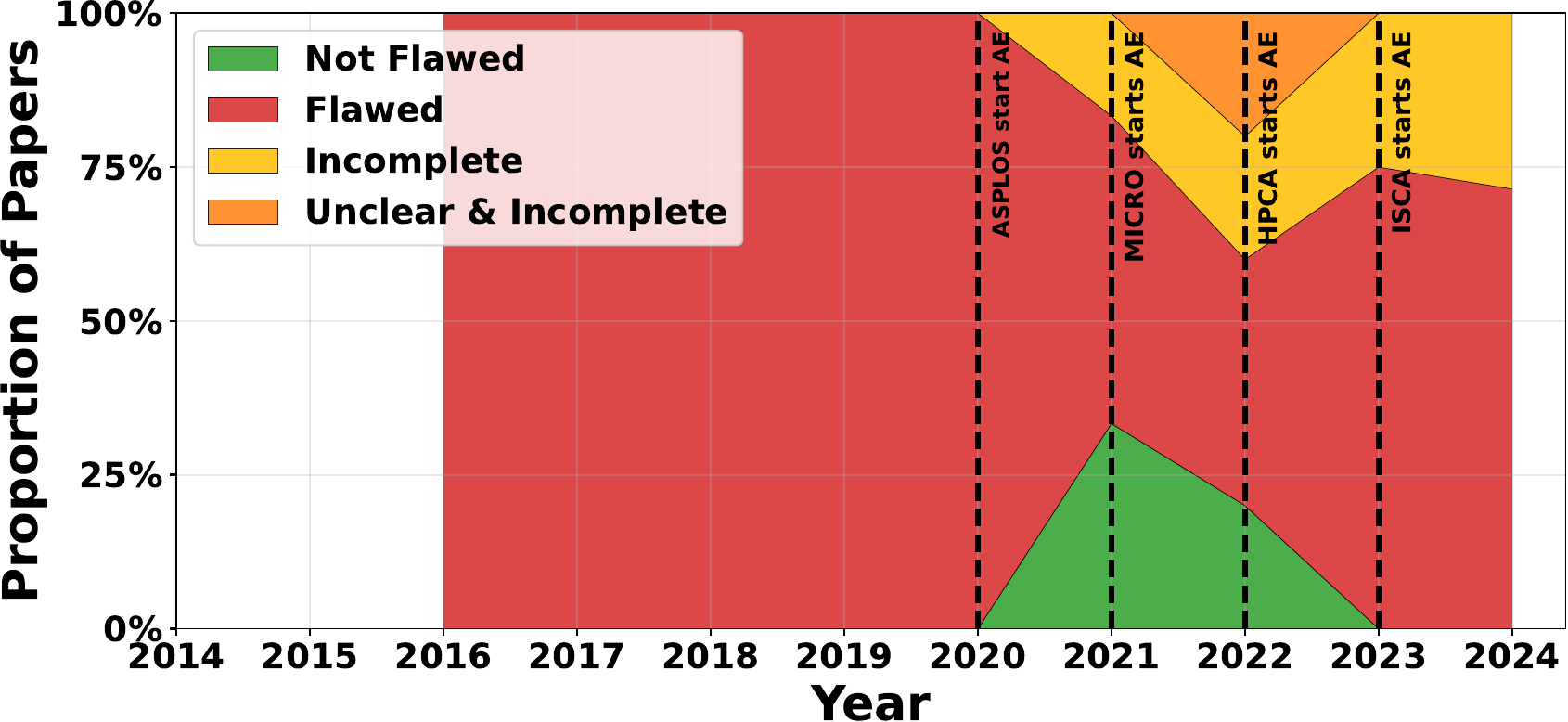}
    \caption{Availability of code, materials, and documentation for papers in architecture conferences over time (with AE start).}
    \label{fig:porportion_plot_C3_architecture}
\end{figure}

\begin{table*}[tb]
    \scriptsize
    \caption{Papers selected for this study, sorted by year, venue, and title.}
    \begin{threeparttable}[h]
    \begin{adjustbox}{center}
        \begin{tabular}{clll}
            \toprule
            Year          & Venue    &  Authors                 & Title \\
            \midrule
            2014          & USENIX   & Yarom and Falkner~\cite{yarom2014flush}        & FLUSH+RELOAD: a High Resolution, Low Noise, L3 Cache Side-Channel Attack \\
            2015          & CCS      & Oren et al.~\cite{oren2015spy}             & The Spy in the Sandbox: Practical Cache Attacks in JavaScript and their Implications \\
            2015          & S\&P     & Liu et al.~\cite{liu2015lastlevel}              & Last-Level Cache Side-Channel Attacks are Practical \\
            2015          & S\&P     & Irazoqui et al.~\cite{irazoquiapecechea2015s$a}         & S\$A: A Shared Cache Attack that Works Across Cores and Defies VM Sandboxing—and its Application to AES \\
            2015          & USENIX   & Gruss et al.~\cite{gruss2015cache}            & Cache Template Attacks: Automating Attacks on Inclusive Last-Level Caches \\
            2016          & CCS      & Evtyushkin et al.~\cite{evtyushkin2016covert}       & Covert Channels through Random Number Generator: Mechanisms, Capacity Estimation and Mitigations \\
            2016          & MICRO    & Evtyushkin et al.~\cite{evtyushkin2016jump}       & Jump over ASLR: Attacking branch predictors to bypass ASLR \\
            2016          & USENIX   & Lipp et al.~\cite{lipp2016armageddon}             & ARMageddon: Cache Attacks on Mobile Devices \\
            2016          & USENIX   & Pessl et al.~\cite{pessl2016drama}            & DRAMA: Exploiting DRAM Addressing for Cross-CPU Attacks \\
            2017          & MICRO    & Naghibijouybari et al.~\cite{naghibijouybari2017constructing}  & Constructing and Characterizing Covert Channels on GPGPUs \\
            2017          & NDSS     & Gras et al.~\cite{gras2017aslr}             & ASLR on the Line: Practical Cache Attacks on the MMU \\
            2017          & NDSS     & Maurice et al.~\cite{maurice2017hello}          & Hello from the Other Side: SSH over Robust Cache Covert Channels in the Cloud \\
            2017          & USENIX   & Disselkoen et al.~\cite{disselkoen2017prime}       & Prime+Abort: A Timer-Free High-Precision L3 Cache Attack using Intel TSX \\
            2017          & USENIX   & Bulck et al.~\cite{bulck2017telling}            & Telling Your Secrets Without Page Faults: Stealthy Page Table-Based Attacks on Enclaved Execution \\
            2018          & ASPLOS   & Evtyushkin et al.~\cite{evtyushkin2018branchscope}      & BranchScope: A New Side-Channel Attack on Directional Branch Predictor \\
            2018          & CCS      & Naghibijouybari et al.~\cite{naghibijouybari2018rendered}  & Rendered Insecure: GPU Side Channel Attacks are Practical \\
            2018          & CCS      & Shin et al.~\cite{shin2018unveiling}             & Unveiling Hardware-based Data Prefetcher, a Hidden Source of Information Leakage \\
            2018          & HPCA     & Yao et al.~\cite{yao2018are}              & Are Coherence Protocol States Vulnerable to Information Leakage? \\
            2018          & S\&P     & Frigo et al.~\cite{frigo2018grand}            & Grand Pwning Unit: Accelerating Microarchitectural Attacks with the GPU \\
            2018          & USENIX   & Schaik et al.~\cite{schaik2018malicious}           & Malicious Management Unit: Why Stopping Cache Attacks in Software is Harder Than You Think \\
            2018          & USENIX   & Gras et al.~\cite{gras2018translation}             & Translation Leak-aside Buffer: Defeating Cache Side-channel Protections with TLB Attacks \\
            2019          & HPCA     & Khatamifard et al.~\cite{khatamifard2019powert}      & POWERT Channels: A Novel Class of Covert Communication Exploiting Power Management Vulnerabilities \\
            2019          & S\&P     & Yan et al.~\cite{yan2019attack}              & Attack Directories, Not Caches: Side-Channel Attacks in a Non-Inclusive World \\
            2019          & S\&P     & Aldaya et al.~\cite{aldaya2019port}           & Port Contention for Fun and Profit \\
            2019          & S\&P     & Vila et al.~\cite{vila2019theory}             & Theory and Practice of Finding Eviction Sets \\
            2020          & HPCA     & Xiong et al.~\cite{xiong2020leaking}            & Leaking Information Through Cache LRU States \\
            2020          & USENIX   & Briongos et al.~\cite{briongos2020reload}         & RELOAD+REFRESH: Abusing Cache Replacement Policies to Perform Stealthy Cache Attacks \\
            2021          & ASPLOS   & Saileshwar et al.~\cite{saileshwar2021streamline}        & Streamline: A Fast, Flushless Cache Covert-Channel Attack by Enabling Asynchronous Collusion \\
            2021          & CCS      & Purnal et al.~\cite{purnal2021prime}           & Prime+Scope: Overcoming the Observer Effect for High-Precision Cache Contention Attacks \\
            2021          & S\&P     & Tan et al.~\cite{tan2021invisible}              & Invisible Probe: Timing Attacks with PCIe Congestion Side-channel \\
            2021          & S\&P     & Lipp et al.~\cite{lipp2021platypus}             & PLATYPUS: Software-based Power Side-Channel Attacks on x86 \\
            2021          & ISCA     & Ren et al.~\cite{ren2021see}              & I See Dead µops: Leaking Secrets via Intel/AMD Micro-Op Caches \\
            2021          & ISCA     & Haj et al.~\cite{haj-yahya2021ichannels}              & IChannels: Exploiting Current Management Mechanisms to Create Covert Channels in Modern Processors \\
            2021          & ISCA     & Dutta et al.~\cite{dutta2021leaky}            & Leaky Buddies: Cross-Component Covert Channels on Integrated CPU-GPU Systems \\
            2021          & MICRO    & Ahn et al.~\cite{ahn2021networkonchip}              & Network-on-Chip Microarchitecture-based Covert Channel in GPUs \\
            2021          & MICRO    & Kim et al.~\cite{kim2021uccheck}              & UC-Check: Characterizing Micro-operation Caches in x86 Processors and Implications in Security and Performance \\
            2021          & USENIX   & Paccagnella et al.~\cite{paccagnella2021lord}      & Lord of the Ring(s): Side Channel Attacks on the CPU On-Chip Ring Interconnect Are Practical \\
            2022          & ASPLOS   & Yang et al.~\cite{yang2022eavesdropping}            & Eavesdropping User Credentials via GPU Side Channels on Smartphones \\
            2022          & HPCA     & Cui et al.~\cite{cui2022abusing}              & Abusing Cache Line Dirty States to Leak Information in Commercial Processors \\
            2022          & HPCA     & Kim et al.~\cite{kim2022dprime}              & DPrime+DAbort: A High-Precision and Timer-Free Directory-Based Side-Channel Attack in Non-Inclusive Cache Hierarchies using Intel TSX \\
            2022          & HPCA     & Deng et al.~\cite{deng2022leaky}             & Leaky Frontends: Security Vulnerabilities in Processor Frontends \\
            2022          & S\&P     & Guo et al.~\cite{guo2022adversarial}              & Adversarial Prefetch: New Cross-Core Cache Side Channel Attacks \\
            2022          & S\&P     & Vicarte et al.~\cite{vicarte2022augury}          & Augury: Using Data Memory-Dependent Prefetchers to Leak Data at Rest \\
            2022          & S\&P     & Wan et al.~\cite{wan2022meshup}              & MeshUp: Stateless Cache Side-channel Attack on CPU Mesh \\
            2022          & MICRO    & Guo et al.~\cite{guo2022leaky}              & Leaky Way: A Conflict-Based Cache Covert Channel Bypassing Set Associativity \\
            2022          & USENIX   & Lipp et al.~\cite{lipp2022amd}             & AMD Prefetch Attacks through Power and Time \\
            2022          & USENIX   & Zhao et al.~\cite{zhao2022binoculars}             & Binoculars: Contention-Based Side-Channel Attacks Exploiting the Page Walker \\
            2022          & USENIX   & Dai et al.~\cite{dai2022dont}              & Don’t Mesh Around: Side-Channel Attacks and Mitigations on Mesh Interconnects \\
            2022          & USENIX   & Purnal et al.~\cite{purnal2022double}           & Double Trouble: Combined Heterogeneous Attacks on Non-Inclusive Cache Hierarchies \\
            2022          & USENIX   & Wang et al.~\cite{wang2022hertzbleed}             & Hertzbleed: Turning Power Side-Channel Attacks Into Remote Timing Attacks on x86 \\
            2022          & USENIX   & Aldaya et al.~\cite{aldaya2022hyperdegrade}           & HyperDegrade: From GHz to MHz Effective CPU Frequencies \\
            2022          & USENIX   & Tatar et al.~\cite{tatar2022tlbdr}            & TLB;DR: Enhancing TLB-based Attacks with TLB Desynchronized Reverse Engineering \\
            2023          & ASPLOS   & Chen et al.~\cite{chen2023afterimage}            & AfterImage: Leaking Control Flow Data and Tracking Load Operations via the Hardware Prefetcher \\
            2023          & CCS      & Zhang et al.~\cite{zhang2023tunnels}            & TunneLs for Bootlegging: Fully Reverse-Engineering GPU TLBs for Challenging Isolation Guarantees of NVIDIA MIG \\
            2023          & S\&P     & Gerlach et al.~\cite{gerlach2023security}          & A Security RISC: Microarchitectural Attacks on Hardware RISC-V CPUs \\
            2023          & S\&P     & Kim et al.~\cite{kim2023devious}              & DevIOus: Device-Driven Side-Channel Attacks on the IOMMU \\
            2023          & S\&P     & Wang et al.~\cite{wang2023dvfs}             & DVFS Frequently Leaks Secrets: Hertzbleed Attacks Beyond SIKE, Cryptography, and CPU-Only Data \\
            2023          & S\&P     & Gast et al.~\cite{gast2023squip}             & SQUIP: Exploiting the Scheduler Queue Contention Side Channel \\
            2023          & ISCA     & Yu et al.~\cite{yu2023all}               & All Your PC Are Belong to Us: Exploiting Non-control-Transfer Instruction BTB Updates for Dynamic PC Extraction \\
            2023          & ISCA     & Dutta et al.~\cite{dutta2023spy}            & Spy in the GPU-box: Covert and Side Channel Attacks on Multi-GPU Systems \\
            2023          & MICRO    & Guo et al.~\cite{guo2023uncore}              & Uncore Encore: Covert Channels Exploiting Uncore Frequency Scaling \\
            2023          & USENIX   & Zhang et al.~\cite{zhang2023mwait}            & (M)WAIT for It: Bridging the Gap between Microarchitectural and Architectural Side Channels \\
            2023          & USENIX   & Zhang et al.~\cite{zhang2023bunnyhop}            & BunnyHop: Exploiting the Instruction Prefetcher \\
            2023          & USENIX   & Kogler et al.~\cite{kogler2023collide}           & Collide+Power: Leaking Inaccessible Data with Software-based Power Side Channels \\
            2023          & USENIX   & Wang et al.~\cite{wang2023nvleak}             & NVLeak: Off-Chip Side-Channel Attacks via Non-Volatile Memory Systems \\
            2023          & USENIX   & Liu et al.~\cite{liu2023sidechannel}              & Side-Channel Attacks on Optane Persistent Memory \\
            2023          & USENIX   & Yu et al.~\cite{yu2023synchronization}               & Synchronization Storage Channels (S2C): Timer-less Cache Side-Channel Attacks on the Apple M1 via Hardware Synchronization Instructions \\
            2023          & USENIX   & Katzman et al.~\cite{katzman2023gates}          & The Gates of Time: Improving Cache Attacks with Transient Execution \\
            2024          & ASPLOS   & Yavarzadeh et al.~\cite{yavarzadeh2024pathfinder}              & Pathfinder: High-Resolution Control-Flow Attacks Exploiting the Conditional Branch Predictor \\
            2024          & ASPLOS   & Zhao et al.~\cite{zhao2024last}          & Last-Level Cache Side-Channel Attacks are Feasible in the Modern Public Cloud \\
            2024          & CCS      & Rauscher and Gruss~\cite{rauscher2024crosscore}       & Cross-Core Interrupt Detection: Exploiting User and Virtualized IPIs \\
            2024          & CCS      & Wilke et al.~\cite{wilke2024tdxdown}            & TDXdown: Single-Stepping and Instruction Counting Attacks against Intel TDX \\
            2024          & HPCA     & Xu et al.~\cite{xu2024exploitation}               & Exploitation of Security Vulnerability on Retirement \\
            2024          & HPCA     & Chen et al.~\cite{chen2024prefetchx}             & PREFETCHX: Cross-Core Cache-Agnostic Prefetcher-based Side-Channel Attacks \\
            2024          & S\&P     & Rodrigues et al.~\cite{rodrigues2024busted}        & BUSted!!! Microarchitectural Side-Channel Attacks on the MCU Bus Interconnect \\
            2024          & S\&P     & Wang et al.~\cite{wang2024gpuzip}             & GPU.zip: On the Side-Channel Implications of Hardware-Based Graphical Data Compression \\
            2024          & ISCA     & Chowdhuryy et al.~\cite{chowdhuryy2024metaleak}       & MetaLeak: Uncovering Side Channels in Secure Processor Architectures Exploiting Metadata \\
            2024          & MICRO    & Jin et al.~\cite{jin2024ghost}              & Ghost Arbitration: Mitigating Interconnect Side-Channel Timing Attacks in GPU \\
            2024          & MICRO    & Miao et al.~\cite{miao2024veiled}             & Veiled Pathways: Investigating Covert and Side Channels Within GPU Uncore \\
            2024          & NDSS     & Rauscher et al.~\cite{rauscher2024idleleak}         & IdleLeak: Exploiting Idle State Side Effects for Information Leakage \\
            2024          & USENIX   & Chen et al.~\cite{chen2024gofetch}             & GoFetch: Breaking Constant-Time Cryptographic Implementations Using Data Memory-Dependent Prefetchers \\
            2024          & USENIX   & Zhang et al.~\cite{zhang2024invalidate}            & Invalidate+Compare: A Timer-Free GPU Cache Attack Primitive \\
            2024          & USENIX   & O'Connell et al.~\cite{oconnell2024pixel}        & Pixel Thief: Exploiting SVG Filter Leakage in Firefox and Chrome \\
        \end{tabular}
    \end{adjustbox}
    \end{threeparttable}
    \label{tab:papers}
\end{table*}

\end{document}